\documentclass[
reprint,
superscriptaddress,
amsmath,amssymb,
showkeys,
aps,
prb,
]{revtex4-2}

\usepackage{threeparttable}
\usepackage{setspace}
\usepackage{makecell}
\usepackage{amsmath}
\usepackage{graphicx}
\usepackage{dcolumn}
\usepackage{bm}
\usepackage{hyperref}
\usepackage[mathlines]{lineno}
\usepackage{textcomp}
\usepackage{gensymb}
\usepackage{subfigure}
\usepackage{url}
\usepackage{xcolor}
\usepackage{hhline}
\usepackage[version=3]{mhchem}
\usepackage{multirow}
\usepackage[american]{babel}
\usepackage{hhline}
\usepackage{textgreek}
\usepackage{miller}
\usepackage{lipsum}

\definecolor{red}{rgb}{0.0,0.0,0.0}\def\red{\color{red}}

\definecolor{green}{rgb}{0.1,0.6,0.1}
\definecolor{yellow}{rgb}{0.8,0.8,0.6}

\definecolor{blue}{rgb}{0.1,0.1,0.6}

\begin{document}
\title{
Orientation-dependent surface radiation damage in \texorpdfstring{\textit{\textbeta}-\ce{Ga2O3}}{} explored by multiscale atomic simulations}

\author{Taiqiao Liu}
\affiliation{The Institute of Technological Sciences, Wuhan University, Wuhan, 430072, China}

\author{Zeyuan Li}
\affiliation{School of Power and Mechanical Engineering, Wuhan University, Wuhan 430072, China}

\author{Junlei Zhao} 
\email{zhaojl@sustech.edu.cn}
\affiliation{Department of Electronic and Electrical Engineering, Southern University of Science and Technology, Shenzhen 518055, China}

\author{Xiaoyu Fei}
\affiliation{The Institute of Technological Sciences, Wuhan University, Wuhan, 430072, China}

\author{Jiaren Feng}
\affiliation{The Institute of Technological Sciences, Wuhan University, Wuhan, 430072, China}

\author{Yijing Zuo}
\affiliation{The Institute of Technological Sciences, Wuhan University, Wuhan, 430072, China}

\author{Mengyuan Hua} 
\affiliation{Department of Electronic and Electrical Engineering, Southern University of Science and Technology, Shenzhen 518055, China}

\author{Yuzheng Guo}
\affiliation{The Institute of Technological Sciences, Wuhan University, Wuhan, 430072, China}
\affiliation{School of Power and Mechanical Engineering, Wuhan University, Wuhan 430072, China}

\author{Sheng Liu}
\affiliation{The Institute of Technological Sciences, Wuhan University, Wuhan, 430072, China}
\affiliation{School of Power and Mechanical Engineering, Wuhan University, Wuhan 430072, China}

\author{Zhaofu Zhang}
\email{zhaofuzhang@whu.edu.cn}
\affiliation{The Institute of Technological Sciences, Wuhan University, Wuhan, 430072, China}
\affiliation{Hubei Key Laboratory of Electronic Manufacturing and Packaging Integration, Wuhan University, Wuhan, 430072, China}
\affiliation{Suzhou Institute of Wuhan University, Suzhou, 215123, China}


\begin{abstract}

Ultrawide bandgap semiconductor $\beta$-\ce{Ga2O3} holds extensive potential for applications in high-radiation environments.
One of the primary challenges in its practical application is unveiling the mechanisms of surface irradiation damage under extreme conditions.
In this study, we investigate the orientation-dependent mechanisms of radiation damage on four experimentally relevant $\beta$-\ce{Ga2O3} surface facets, namely, \hkl(100), \hkl(010), \hkl(001), and \hkl(-201), at various temperatures.
We employ a multiscale atomic simulation approach, combining machine-learning-driven molecular dynamics (ML-MD) simulations and density functional theory (DFT) calculations.
The results reveal that Ga vacancies and O interstitials are the predominant defects across all four surfaces, with the formation of many antisite defects $\mathrm{Ga}_\mathrm{O}$ and few $\mathrm{O}_\mathrm{Ga}$ observed.
Among the two Ga sites and three O sites, the vacancy found in the O2 site is dominant, while the interstitials at the Ga1 and O1 sites are more significant.
Interestingly, the \hkl(010) surface exhibits the lowest defect density, owing to its more profound channeling effect leading to a broader spread of defects.
The influence of temperature on surface irradiation damage of $\beta$-\ce{Ga2O3} should be evaluated based on the unique crystal surface characteristics.
{\red Moreover, the formation energy and defect concentration calculated by DFT corroborate the results of the MD simulations.}
Comprehending surface radiation damage at the atomic level is crucial for assessing the radiation tolerance and predicting the performance changes of $\beta$-\ce{Ga2O3}-based device in high-radiation environments.

\end{abstract}

\maketitle
\section{Introduction} 

The $\beta$-phase of gallium oxide ($\beta$-\ce{Ga2O3}) possesses an ultrawide bandgap of $\sim 4.9$ eV, a high breakdown electric field (8 MV/cm), and high radiation hardness~\cite{2017_perspective, 2021JMST, 2024RN}.
Its excellent radiation resistance allows $\beta$-\ce{Ga2O3} to maintain the desirable structure and electrical properties under harsh radiation environments, providing significant applications in aerospace and nuclear fields~\cite{2019RadiationEffectReview, 2022APLion_irradtaion, SFazarov2023universal}.
The incidence of highly energetic particles, such as neutrons, electrons, protons, and ions, on $\beta$-\ce{Ga2O3} can cause primary radiation damage in the form of point defects~\cite{2019RadiationEffectReview, 2022APLion_irradtaion}.

Previous theoretical studies have shown that Ga interstitials ($\mathrm{Ga}_\mathrm{i}$) can serve as shallow donors but with high formation energy~\cite{2017PeterDFTPhysRevB.95.075208, 2023IntrinsicDefects}.
Ga vacancy ($\mathrm{V}_\mathrm{Ga}$) is the main compensation acceptor. 
O vacancy ($\mathrm{V}_\mathrm{O}$) is a deep-level donor with ionization energy greater than 1~eV~\cite{2010varleyGa2O3}, which may also be the reason for deep-level luminescence~\cite{2017Luminescence}.
These defects can introduce defect levels as trap states or recombination centers within the bandgap, leading to severe and uncontrolled degradation of device performances~\cite{2019RadiationEffectReview, 1974RadEf}.
For example, Cojocaru \textit{et al.}~\cite{1974RadEf} first investigated the effects of irradiation defects on the performance of $\beta$-\ce{Ga2O3} and found that $\mathrm{V}_\mathrm{Ga}$ defects resulted in decreased conductivity after irradiation.
Ingebrigtsen \textit{et al.}~\cite{2019proton} irradiated $\beta$-\ce{Ga2O3} thin films and single crystals with 0.6 MeV and 1.9 MeV protons, attributing the reduction in carrier concentration to Ga interstitial and antisite defects that pin the Fermi level 0.5 eV below the conduction band minimum.
Lee \textit{et al.}~\cite{2018Lee} used 1.5 MeV electron irradiation on $\beta$-\ce{Ga2O3} rectifiers and observed that irradiation damage can reduce carrier lifetime by increasing charge capture traps, thereby facilitating the recombination of electrons and holes.
However, surfaces and interfaces are often sinks for radiation-generated mobile defects~\cite{2010Interface-defect, 2021YANGinterface-defect}.
Studies on materials, such as tungsten~\cite{Weiguo_W2018, Weiguo_2019}, diamond~\cite{Liu_2022}, and GaN~\cite{2012_GaNSurfaceDamage}, have demonstrated that various crystal surfaces exhibit distinct sensitivities to radiation damage.
This knowledge provides a basis for selecting appropriate surface orientations to improve the radiation resistance of materials.
Additionally, the structural and physicochemical properties of defects on the crystal surface can also significantly influence its electrical properties, causing the performance degradation and reliability concerns~\cite{2024SchottkyBarrierIrradiation}.
Therefore, the insightful understanding of primary radiation damage on different $\beta$-\ce{Ga2O3} surfaces can facilitate the more effective design of radiation-resistant devices.

The most commonly employed surfaces of $\beta$-\ce{Ga2O3} for epitaxial growth and device applications are \hkl(100), \hkl(010), \hkl(001), and \hkl(-201).
Large-sized crystals can be grown easily on the \hkl(100) surface.
However, the cleave plane makes cracks more likely under mechanical pressure because of the low surface energy, leading to poor quality~\cite{2019Large(100), 2006Low-indexSufaces, MuSai2020(100)}.
The \hkl(010) substrate is a common choice for epitaxy owing to its commercial availability, high quality, and the symmetry of its growth surface, which prevents twin formation compared with other surfaces~\cite{Sasaki_2012(010)surface, 2020_010surface, Tadjer_2021}.
The \hkl(001) surface is conducive to large-scale growth and has high defect formation energy~\cite{WANG-dft2023}.
$\beta$-\ce{Ga2O3} grows well on the \hkl(-201) plane and is extensively used for homoepitaxial growth~\cite{2020homoepitaxial(-201), 2021homoepitaxial(-201)}.
Polyakov \textit{et al.}~\cite{2021protonJAP} found that the crystallographic orientation of $\beta$-\ce{Ga2O3} significantly influences its properties under proton irradiation.
Theoretically, studies have been conducted on the threshold displacement energies (TDEs) in $\beta$-\ce{Ga2O3}~\cite{2023JAP, he2024threshold}.
Although numerous experimental studies were undertaken to investigate how radiation impacts $\beta$-\ce{Ga2O3} and its device, the mechanism of radiation damage mechanism by cascade collision of the four surfaces has not been reported yet.

% Given the properties mentioned above, the defects in $\beta$-\ce{Ga2O3} will seriously impact the material and device performance.
% Especially in radiation environments, experimentally uncovering the microscopic mechanisms under which the type, structure, and quantity of defects, as well as their impact on material electrical attribution, present a significant challenge.
Herein, we employ multiscale atomic simulations, such as machine-learning-driven molecular dynamics (ML-MD) and density functional theory (DFT), to investigate radiation damage mechanisms on four surfaces under different temperatures.
Clarification is provided on the developmental processes of vacancy and interstitial atoms at gallium (Ga) and oxygen (O) sites, along with particular sites of Ga1, Ga2, O1, O2, and O3 in radiation-induced damage.
First-principles calculations based on DFT were further used to confirm the ML-MD findings and examine the electronic structural characteristics of intrinsic defects.
This work can guide the design and manufacturing processes of $\beta$-\ce{Ga2O3} to improve their stability and reliability under harsh conditions.
% A deep understanding of the evolution mechanism of two Ga sites and three O sites under different conditions at a microscopic level, to study the impact of irradiation on semiconductor devices.
% This work deepens the understanding of defect distribution on surface irradiation of $\beta$-\ce{Ga2O3}, providing an important theoretical basis for the application of $\beta$-\ce{Ga2O3}-based power electronic devices in extreme environments.

\section{Methodology} 

\subsection{ML-MD simulation}
\begin{figure}[ht!]
    \centering
    \includegraphics[width=8.6cm]{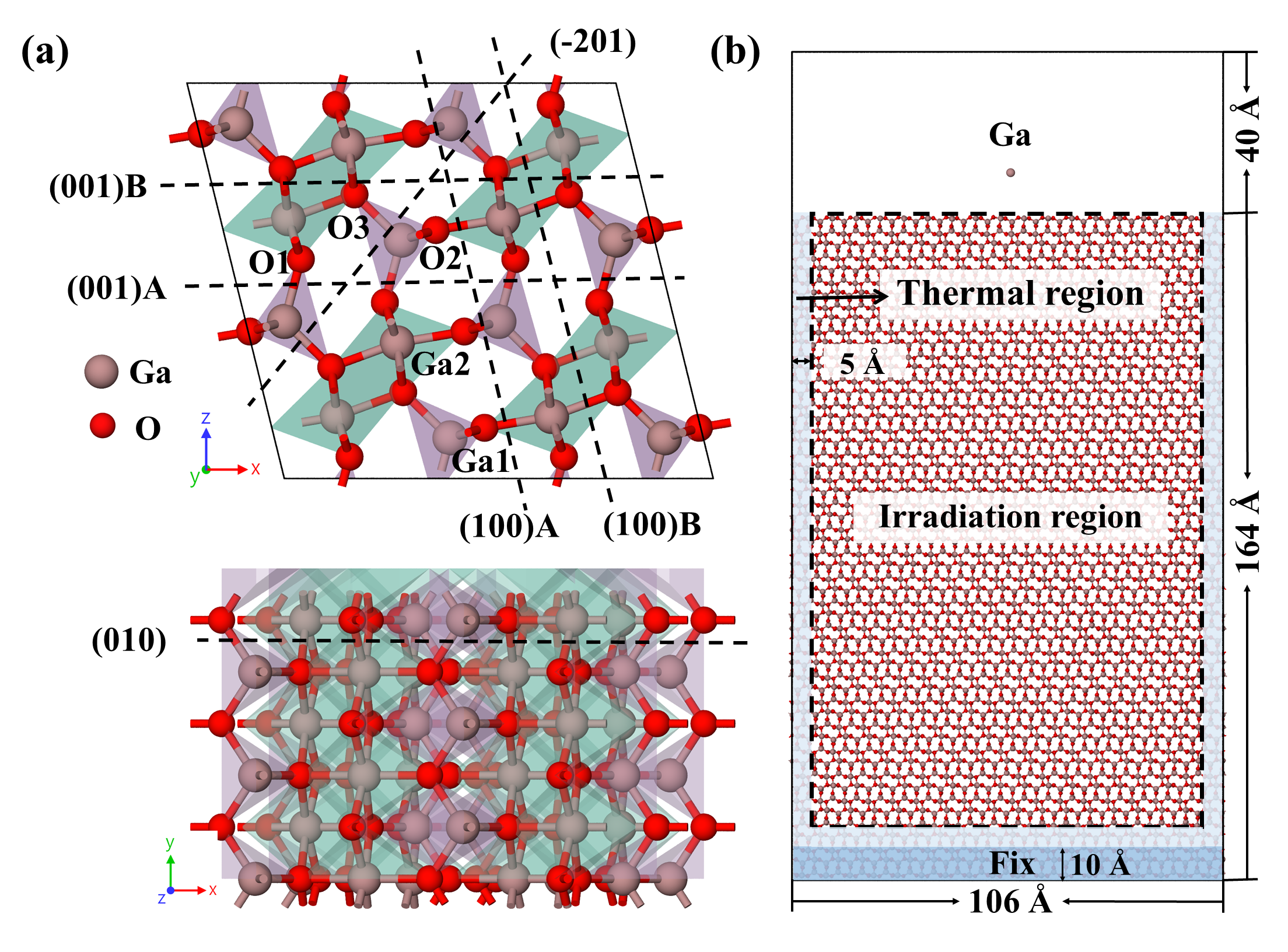}
    \caption{(a) The atomic structure of $\beta$-\ce{Ga2O3} and surface orientations. 
    (b) The side view of surface \hkl(100)B.}
    \label{fig:fig1}
\end{figure}

The large-scale atomic/molecular massively parallel simulator (LAMMPS) code~\cite{lammps2022} and the previously developed \ce{Ga2O3} ML interatomic potential~\cite{SFzhao2023complex} were used for the ML-MD simulations within this work.
Fig.~\ref{fig:fig1}(a) depicts the atomic structure of $\beta$-\ce{Ga2O3} and surface orientations.
Ga has two types of sites: a tetrahedral site (Ga1) and an octahedral site (Ga2). 
O has three types of sites: O1 is threefold coordinated with one Ga1 site and two Ga2 sites, O2 is threefold coordinated with two Ga1 sites and one Ga2 site, and O3 is fourfold coordinated with one Ga1 site and three Ga2 sites.
The B-type terminations of \hkl(100) and \hkl(001) planes are more stable than the corresponding A-type terminations~\cite{WANG-dft2023}.
Their atomic structures from top view are presented in the supplementary material Figure~S1.
Hence, the \hkl(100)B and \hkl(001)B planes are used for surface simulations.
All the \hkl(100) and \hkl(001) surfaces mentioned in this work specifically refer to \hkl(100)B and \hkl(001)B, respectively.
Periodic boundary conditions are utilized in the $x$ and $y$ directions, whereas a free surface is used in the $z$ direction.
The initial models of open surface \hkl(100), \hkl(010), \hkl(001) and \hkl(-201) contain 174,960 ($\sim 106 \times 110 \times 204$ \r A$^{3}$), 174,800 ($\sim 121 \times 112 \times 237$~\r A$^{3}$), 151,470 ($\sim 112 \times 102 \times 193$ \r A$^{3}$) and 190,080 ($\sim 120 \times 111 \times 202$ \r A$^{3}$) atoms, respectively.
A vacuum layer of 40 \r A thickness is adopted.
The primary knock-on atom (PKA) is Ga atom which is 10 \r A away from the \ce{Ga2O3} surface. 

The adopted \hkl(100) surface model setting is shown in Fig.~\ref{fig:fig1}(b).
The other three surfaces are shown in the supplementary material Figure~S2.
As seen in Fig.~\ref{fig:fig1}(b), the temperature was controlled using a Nos{\'e}-Hoover thermostat~\cite{hoover1985} at the marked thermal region of the simulation cell, employing canonical ensemble ($NVT$).
The bottom layer is fixed with 10 \r A-thick \ce{Ga2O3} atoms.
The middle area is the irradiation region with a micro-canonical ensemble ($NVE$).
The irradiation process is conducted using the following methods.
First, each of four surface models is relaxed for 50 ps under the isothermal-isobaric ensemble ($NPT$), followed by 20 ps under $NVT$.
Then, these relaxed models are irradiated for 50 ps with PKA energy of 1.5 keV at temperatures of 173 K, 300 K, and 500 K, respectively.
To avoid channeling effects, the PKA atom is set to an incidence angle deviation of $7\degree$~\cite{Nordlund-PhysRevB.94.214109}.
We apply the adaptive MD timestep in the cascade collision irradiation region to guarantee the maximum moving distance of atoms per MD step is less than 0.1 \r A.
Electron stopping is applied as a friction term to atoms with kinetic energy greater than 10 eV~\cite{nordlund1998defect}.
For statistical analysis, ten independent simulations were performed at different temperatures and surface orientations, so $10 \times 3 \times 4 = 120$ simulations in total.
Based on the cascade collision irradiation results, we analyzed the Ga and O defect configurations under different conditions using Wigner-Seitz (WS) method implemented in the Open Visualization Tool (OVITO) package~\cite{ovito2010}.

\subsection{First-principles calculations}

% Given the limitations of classical ML-MD simulations in accounting for electronic properties, 
To study the electronic properties of $\beta$-\ce{Ga2O3}, first-principles calculations based on DFT are employed to investigate the defect and carrier concentrations of various defect configurations in $\beta$-\ce{Ga2O3}, as implemented.
The DFT calculations are completed by the Vienna Ab initio Simulation Package (VASP) software~\cite{PhysRevB.54.11169}.
The projector-augmented wave pseudopotentials~\cite{PhysRevB.50.17953_PAW} are employed, with a plane-wave energy cutoff of 520 eV.
The Perdew-Burke-Ernzerhof version of the generalized gradient approximations (PBE-GGA)~\cite{PhysRevLett.77.3865} is used for structure relaxation. 
An intrinsic bandgap of 4.86 eV for $\beta$-\ce{Ga2O3} is obtained using the Heyd-Scuseria-Ernzerhof (HSE06) hybrid functional~\cite{2006JChPh.124u9906H_HSE} with a Hartree-Fock fraction of 0.34 throughout this work~\cite{Varley2020_book}.
During structural relaxation, the perfect cell is allowed to fully relax until the atomic force is less than 0.01 eV$\cdot$\r A$^{-1}$.
This process yields conventional-cell lattice constants as $a = 12.27$ \r A, $b = 3.05$ \r A, $c = 5.82$ \r A, and $\angle \beta = 103.7\degree$, perfectly consistent with report~\cite{2010varleyGa2O3}.
Defect calculations were performed using the defect and dopant ab initio simulation package (DASP)~\cite{Huang_2022}.
For the point defect calculations, a 160-atom supercell is used, where only single $\Gamma$ point was considered for Brillouin zone sampling. 
The defect formation energy, $E_{f}[\mathrm{defect},q]$, was given by the formula~\cite{2004vdwreview}:
\begin{equation}\label{eq:1}
\begin{split}
E_{f}[\mathrm{defect},q]    & =  E_{\mathrm{tot}}[\mathrm{defect},q] - E_{\mathrm{tot}}[\mathrm{bulk}] +  \\ 
                            & + \sum_{i} n_{i} \mu_{i} + q(E_\mathrm{Fermi} + E_\mathrm{v}+\Delta V),
\end{split}
\end{equation}
where $E_{\mathrm{tot}}[\mathrm{defect}]$ and $E_{\mathrm{tot}}[\mathrm{bulk}]$ are the calculated total energies of the supercell with a defect in charge state q and the perfect supercell of $\beta$-\ce{Ga2O3}, respectively.
$\mu_{i}$ donates the chemical potential of Ga atoms or O atoms, depending on environmental condition. 
$E_\mathrm{Fermi}$ is the Fermi level referenced to the valence band maximum $E_\mathrm{v}$.
The chemical potential $\mu_\mathrm{O}$ can vary from being in equilibrium with \ce{O2} ($\mu_\mathrm{O} = \frac{1}{2} E_{\mathrm{tot}}[\mathrm{O}_{2}\mathrm{(g)}]$), defining O-rich/Ga-poor conditions, to an extreme of O-poor/Ga-rich limit where it is further constrained by the formation enthalpy of $\beta$-\ce{Ga2O3} ($\mu_\mathrm{O} = \frac{1}{2} E_{\mathrm{tot}}[\mathrm{O}_{2}\mathrm{(g)}] + \frac{1}{3} \Delta H[\mathrm{Ga}_{2}\mathrm{O}_{3}]$). $\mu_{\mathrm{Ga}}$ relates to the energy per atom in elemental Ga in the O-poor/Ga-rich limit with an added contribution of up to $\frac{1}{2} \Delta H[\mathrm{Ga}_{2}\mathrm{O}_{3}] - 3\mu_\mathrm{O}$. 
The $\Delta V$ term defines the finite-size correction for charged defects following the Freysoldt-Neugebauer-Van de Walle (FNV) scheme~\cite{PhysRevLett.102.016402, FNV2010}.

For a defect in its charge state $q$, the equilibrium density $n[\mathrm{defect},q]$ depends on its formation energy, given as~\cite{2021HuangSmall}:
\begin{equation} \label{eq:yourlabel}
n[\mathrm{defect},q] = N_\mathrm{sites} \cdot g_{q} \exp{(-\frac{\Delta E_\mathrm{Fermi}[\mathrm{defect},q]}{k_{B} T})},
\end{equation}
where $n[\mathrm{defect},q]$ is the defect formation energy, $k_\mathrm{B}$ is the Boltzmann constant, $T$ is the temperature, $N_\mathrm{sites}$ is the density of the atomic sites on which the defect level with the defect can form, and $g_{q}$ is the degeneracy factor~\cite{PhysRevB.83.245207_degeneracy_factor}.
The density of hole and electron carriers follows the Boltzmann distribution, given by~\cite{2021HuangSmall}:
\begin{equation}\label{eq:2}
 \begin{cases}
 p_{0} = N_\mathrm{v} \exp{[-E_\mathrm{Fermi}/(k_\mathrm{B} T)]} \\
 n_{0} = N_\mathrm{c} \exp{[(E_\mathrm{Fermi} - E_\mathrm{g})/(k_\mathrm{B} T)]}    
 \end{cases},
\end{equation}
where $N_\mathrm{v}$ and $N_\mathrm{c}$ are the effective density of states for valance band and conduction band edges, respectively.

\section{Results and discussion}

\subsection{ML-MD simulations of defect generation and evolution}

\begin{figure*}[!ht]
    \centering
    \includegraphics[width=15cm]{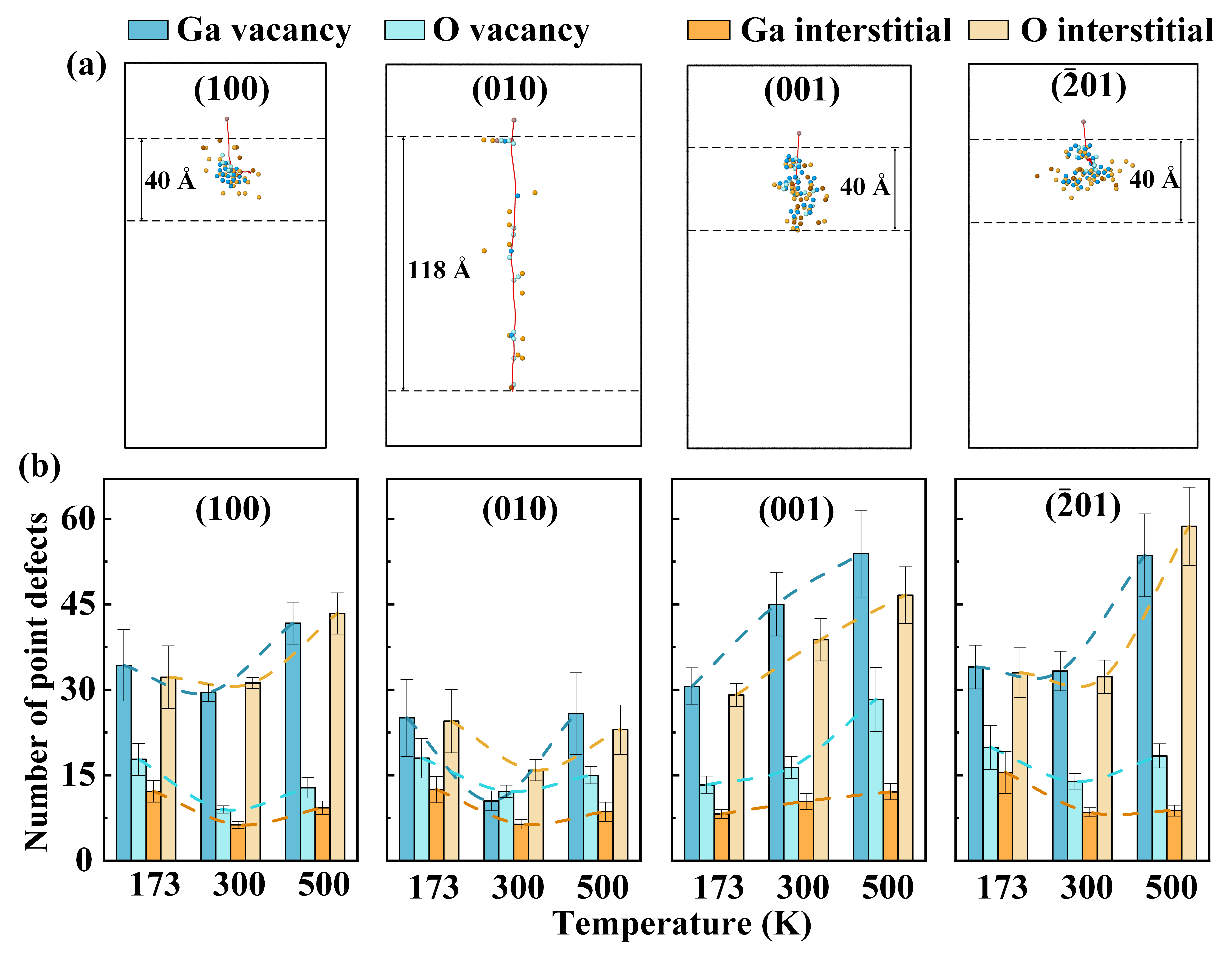}
    \caption{(a) The orientation-dependent defect distribution of $\beta$-\ce{Ga2O3} after cascading collision with 1.5 keV irradiation energy at 300 K. 
    Results of 173 K and 500 K are included in the supplementary material Figure~S3.
    The red line represents ion motion trajectory.
    (b) The statistic number of point defects for different surfaces at 173 K, 300 K and 500 K, respectively.
    {\red The error bars represent the standard errors.}}
    \label{fig:fig2}
\end{figure*}

Fig.~\ref{fig:fig2}(a) shows the point defect (including vacancy and interstitial) distribution of $\beta$-\ce{Ga2O3} after the collision cascade on four surfaces at 300 K.
The other two temperatures are provided in the supplementary material Figure~S3.
All the energies for irradiation are set to 1.5 keV herein.
The supplementary material Figure~S4 displays a top view of radiation damage on the surface of $\beta$-\ce{Ga2O3}.
The atomic arrangements of the four surfaces, with both side and top views, are presented in the supplementary material Figure~S5.
The animations of the four surfaces of $\beta$-\ce{Ga2O3} at 300 K are provided in the Supplementary Video.
Significant axial channeling occurs in \hkl(010) plane.
On the plane, these channels are essentially rows of atoms that form pathways with lower atomic density.
The probability of ion collision is lower in the direction, thereby enhancing the penetration depth.
Consequently, the defect distribution of \hkl(010) surface in Fig.~\ref{fig:fig2} illustrates that despite setting the deviation angle of $7\degree$ to prevent the extensive channeling effect, the incident atom still penetrates deeply beneath the surface, causing defects distributing along the ion track (as indicated by the red line in Fig.~\ref{fig:fig2}(a)).
In contrast, the defects in the other three surfaces are relatively shallow, concentrated within 40 \r A from the surface.

The average number of vacancies and interstitials for both Ga and O is displayed in Fig.~\ref{fig:fig2}(b).
The orientation-dependency of defect quantities on surfaces is illustrated in the supplementary material Figure~S6.
The distinct anisotropy of $\beta$-\ce{Ga2O3} leads to varied properties of the four surface orientations during epitaxial growth.
Our results show that these four crystal surfaces exhibit a varied response to irradiation.
The \hkl(100) surface exhibits minimal defects at 300 K, with the highest defects at 500 K, and 173 K succeeding them.
At 173 K and 500 K, the number of defects on the \hkl(010) surface is nearly constant, while at 300 K, the fewest number of defects is observed. 
The \hkl(010) surface exhibits the least Ga vacancies and O interstitials at the three temperatures among all surfaces.
The Ga interstitial remains the lowest at 300 K and 500 K, while the O vacancy maintains a moderate level.
% suggesting that this particular plane possesses a stable radiation resistance and aligns with the highest quality crystal face achievable in experiments.
However, as shown in Fig.~\ref{fig:fig2}(a), the \hkl(010) surface tends to develop large-depth defects owing to its unique projected atomic arrangement and strong channeling effect.
When considering the utilization of this orientation for device fabrication, it is crucial to carefully consider the impact of the distribution and concentration of point defects.

As the temperature increases, defects gradually increase on the \hkl(001) surface.
For \hkl(-201) surface, there is not much difference between 173 K and 300 K, but at 500 K, the Ga vacancies and O interstitials increase.
Therefore, the radiation resistance of the \hkl(010), \hkl(001) and \hkl(-201) crystallographic planes under the higher temperature (500 K) is worse than that observed at lower (173 K) and constant temperatures (300 K).
%At 173 K, the differences in defects among the four surfaces are minimal.
%At 300 K and 500 K, the (010) plane diverges significantly from the other three planes regarding defect characteristics.
Moreover, \hkl(001) and \hkl(-201) surfaces demonstrate a greater prevalence of defects when compared to the \hkl(100) surface at 500 K.

\begin{figure*}[ht!]
    \centering
    \includegraphics[width=15cm]{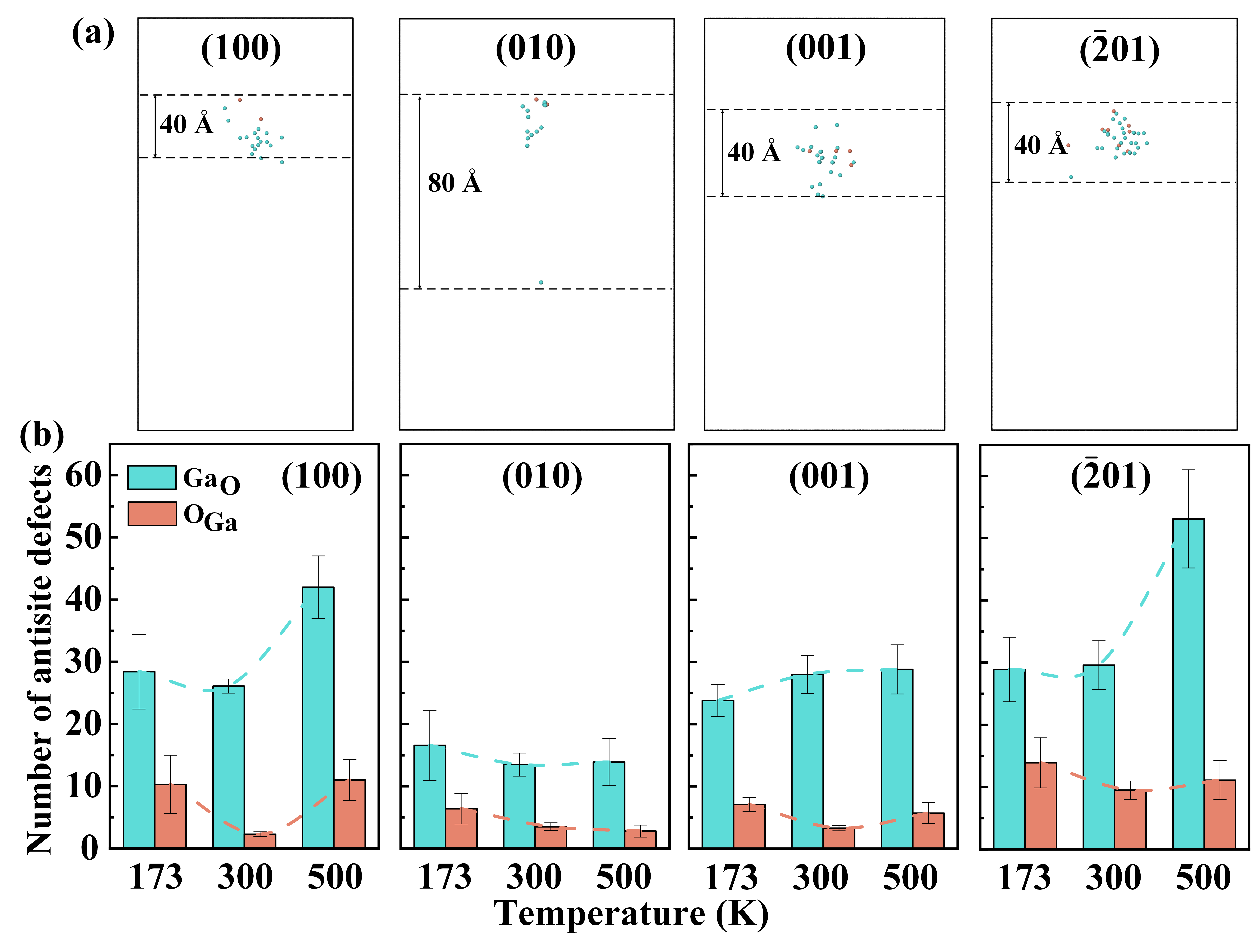}
    \caption{(a) The orientation-dependent antisite defect distribution of $\beta$-\ce{Ga2O3} after cascading at 300 K.
    Results of 173 K and 500 K are included in the supplementary material Figure~S6. 
    (b) The statistic number of antisite defects for different surfaces at 173 K, 300 K and 500 K, respectively.
    {\red The error bars represent the standard deviations.}}
    \label{fig:fig3}
\end{figure*}

At low temperatures, atomic dynamics are reduced in materials, making it simpler for irradiation-induced defects to be preserved.
High temperatures can make point defects generated by irradiation migrate easily and recombine for the dynamic annealing process.
Thus, these defects can be restrained by high temperatures~\cite{2019highT}.
However, the above results suggest that the theory does not apply to $\beta$-\ce{Ga2O3}.
This is the result of several synergistic reasons.
Firstly, as the temperature rises, there is a gradual reduction in TDE, leading to a noticeable statistical rise in defect counts due to a single irradiation event.
Secondly, the automatic healing of the lattice, accelerated by a temperature increase, statistically reduces the defect count.
Moreover, the emergence of more stable and less easily repairing or eliminating defects, such as antisite defects in Fig.~\ref{fig:fig3}, leads to defect accumulation.
Consequently, the overall synergistic effect remains dependent on the specific crystal surface.

\begin{figure*}[ht!]
    \centering
    \includegraphics[width=15cm]{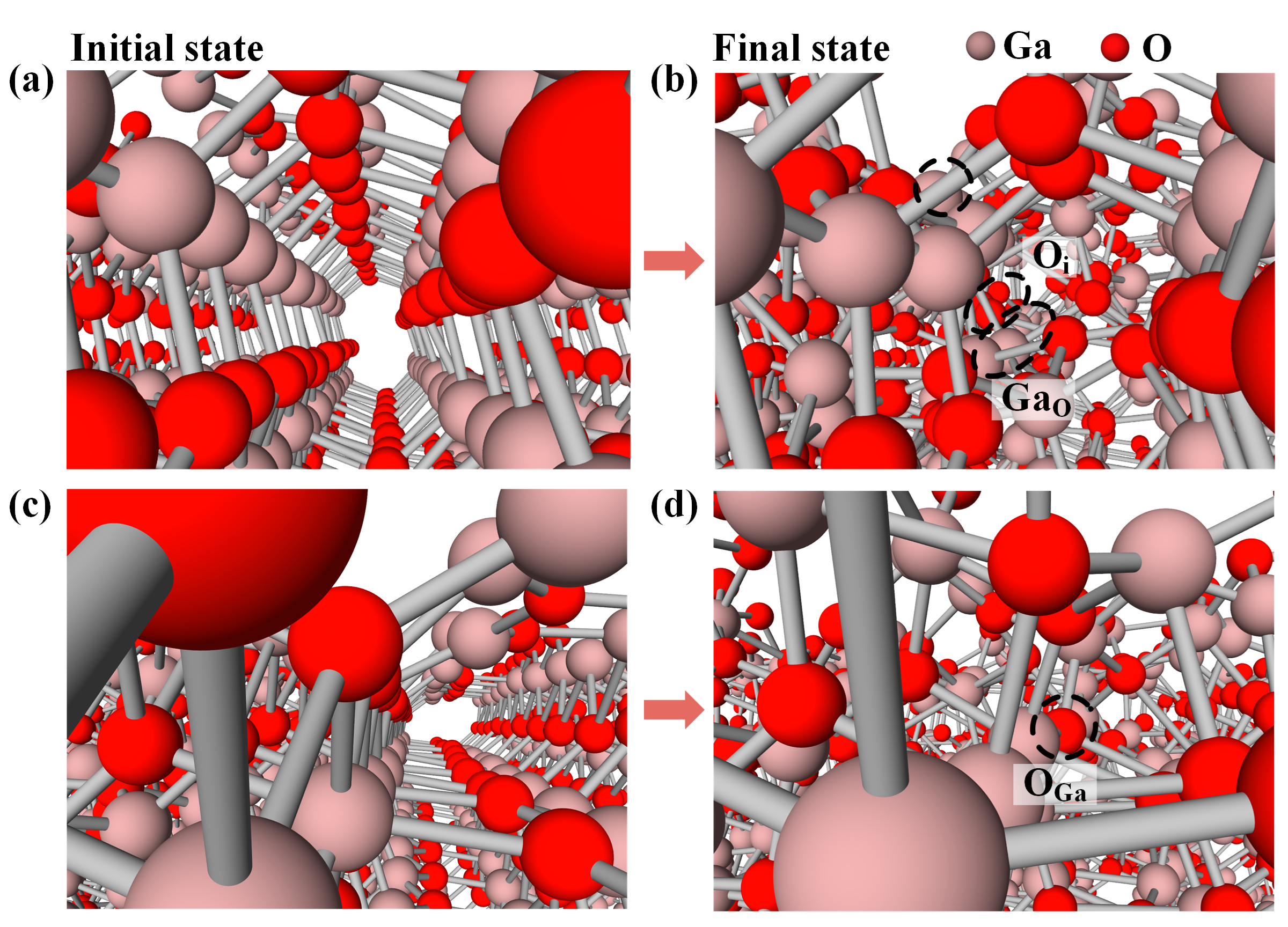}
    \caption{Snapshots of the antisite defect $\mathrm{Ga}_\mathrm{O}$ (b) and $\mathrm{O}_\mathrm{Ga}$ formation from (a, c) the initial states to (b, d) the final states during cascade collision irradiation.}
    \label{fig:fig4}
\end{figure*}

As shown in Fig.~\ref{fig:fig2}(b) and the irradiation studies of $\beta$-\ce{Ga2O3} bulk materials~\cite{SFazarov2023universal, SFzhao2024crystallization}, it is a notable phenomenon that the quantity of O vacancies and Ga interstitials is markedly lower compared with Ga vacancies and O interstitials.
The O interstitial has a higher number of defects because there are more O atoms than Ga atoms in $\beta$-\ce{Ga2O3}.
Thus, there is a higher probability of defects generated by the collision cascade.
The TDE in $\beta$-\ce{Ga2O3} indicates that Ga atoms have greater TDEs than O atoms~\cite{he2024threshold, 2023JAP}.
Additionally, it also relates to antisite defect formation.
Figs.~\ref{fig:fig3}(a) and (c) show the initial state of surface structures that form the antisite $\mathrm{Ga}_\mathrm{O}$ (Fig.~\ref{fig:fig3}(b)) and $\mathrm{O}_\mathrm{Ga}$ (Fig.~\ref{fig:fig3}(d)), respectively.
After the cascade collision, the Ga interstitials recombine with the O sites, and the O atoms are displaced to form interstitials in the final state (Fig.~\ref{fig:fig3}(b)).
Ga atoms are easily recombined, but they preferentially occupy O vacancies instead of Ga vacancies.
This preferential recombination hinders the ability of O atoms to return to their original vacancies, thereby prompting the formation of antisite defects in $\mathrm{Ga}_\mathrm{O}$.
Similarly, a few O interstitials recombine with Ga vacancies to form $\mathrm{O}_\mathrm{Ga}$ as shown in Fig.~\ref{fig:fig3}(b).
The spatial distribution at 300 K and statistics of antisite defects of $\mathrm{Ga}_\mathrm{O}$ and $\mathrm{O}_\mathrm{Ga}$ under different conditions are demonstrated in Fig.~\ref{fig:fig4}.
The supplementary material Figure~S6 illustrates the distribution of antisite defects at temperatures of 173~K and 500~K.
The distribution of antisite defects is near the vacancies and interstitials generated by irradiation.
The distribution depth of antisite defects is comparable to that of interstitials and vacancies within 40 \r A, as illustrated in Fig.~\ref{fig:fig3}(a).
Notably, for the \hkl(010) surface, the distribution depth is significantly shallower.
At the higher temperature of 500~K, a more significant difference in the amounts of Ga interstitials, O vacancies, and the same between Ga vacancies and O interstitials is observed in Fig.~\ref{fig:fig2}(b).
This indicates a higher probability of antisite defects $\mathrm{Ga}_\mathrm{O}$ being formed.
The number of interstitials and vacancies on the \hkl(001) surface at 300 K and 500 K shows minimal difference, similar to the behavior of antisite defects.
Under all conditions, the number of $\mathrm{Ga}_\mathrm{O}$ defects is much greater than the number of $\mathrm{O}_\mathrm{Ga}$ defects, and the result satisfies the relation $\mathrm{V}_\mathrm{O} = \mathrm{O}_\mathrm{i} + \mathrm{O}_\mathrm{Ga} - \mathrm{Ga}_\mathrm{O}$.
It can be explained in more detail in the discussion combined with DFT calculations in Section~\ref{sec:def}.

\subsection{Precise defect types in Ga and O sites}

\begin{figure*}[ht!]
    \centering
    \includegraphics[width=15cm]{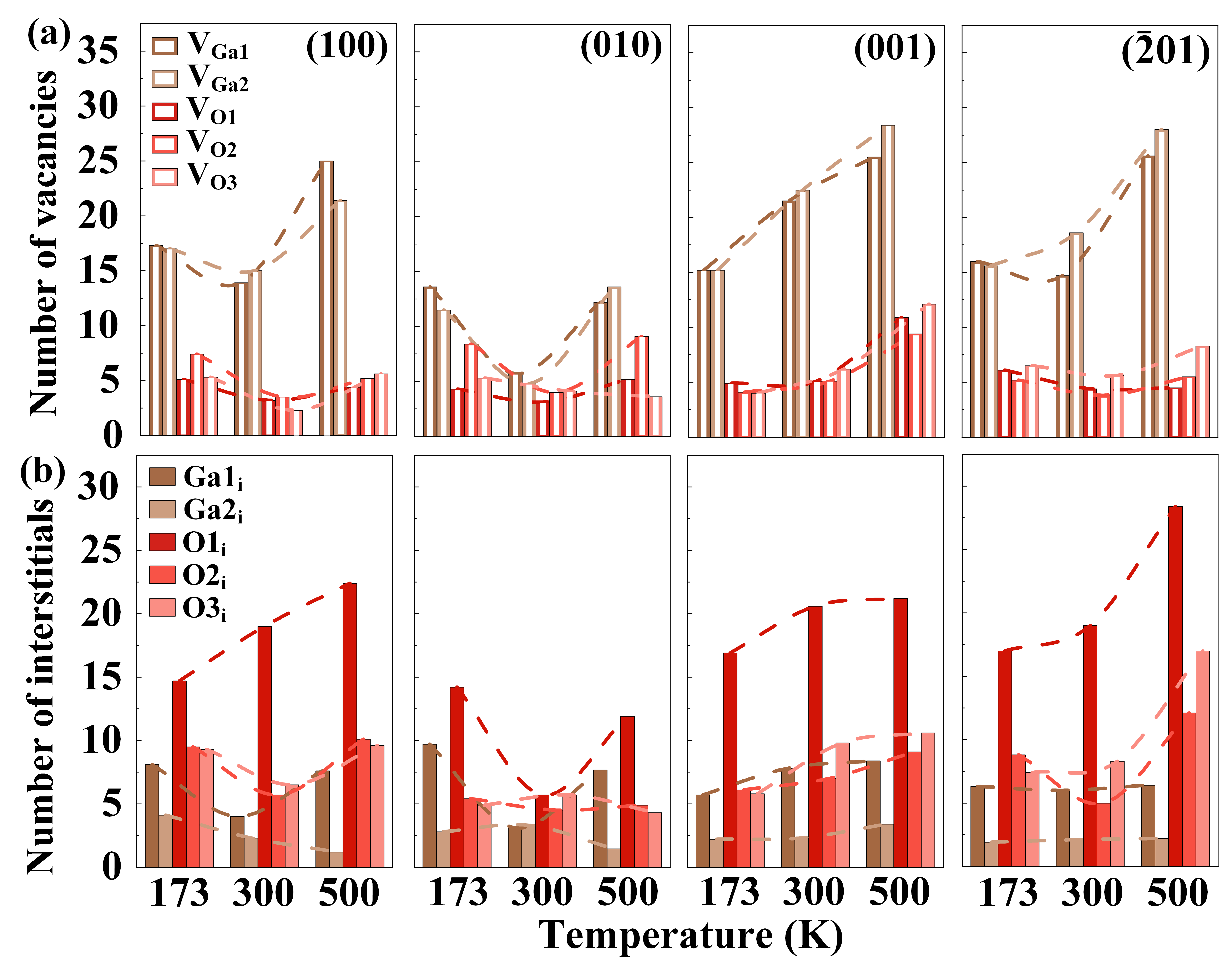}
    \caption{The temperature-dependent defect evolution of (a) vacancy defects and (b) interstitial defects on different surface orientations in $\beta$-\ce{Ga2O3} with 1.5 keV irradiation energy.}
    \label{fig:fig5}
\end{figure*}

To understand the defect evolution in $\beta$-\ce{Ga2O3}, with its complicated crystal structure, it is crucial to understand the precise distribution of Ga1, Ga2, O1, O2, and O3 sites.
We used self-developed Python code to differentiate the defect distribution of various atomic occupancies.
The results are shown in Fig~\ref{fig:fig5}.
The screening criteria are as follows:
(i) Firstly, Ga1 and Ga2 are screened out based on the fact that Ga1 is a four-coordinated atom and Ga2 is a six-coordinated atom;
(ii) Then, based on the results of Ga, O1 is threefold coordinated with 1 Ga1 and 2 Ga2, O2 is threefold coordinated with 2 Ga1 and 1 Ga2, and O3 is threefold coordinated with 1 Ga1 and 3 Ga2 to determine O1, O2, and O3.

As previously noted, the interaction between the Ga interstitials and O vacancies leads to a high occurrence of $\mathrm{Ga}_\mathrm{O}$ defects while the $\mathrm{O}_\mathrm{Ga}$ defects are rare.
Furthermore, upon the formation of the $\mathrm{Ga}_\mathrm{O}$ defects, the oxygen atom cannot rejoin in the O vacancy, making it displaced from the process.
Thus, the results of Ga and O vacancies and TDEs should be employed to decide the displacing ability of Ga and O atoms.
As shown in Fig.~\ref{fig:fig5}, the disparity in vacancy numbers at Ga1 and Ga2 sites for Ga atoms is minor under different conditions. The average TDEs for Ga1 and Ga2 were reported as 22.9 eV and 20.0 eV, respectively~\cite{he2024threshold}.
%However, the count of interstitial atoms varies significantly in most cases, with notably fewer interstitial atoms at the Ga2 site than Ga1.
The lower TDE indicates that the atom is more easily displaced from its lattice site.
In Fig.~\ref{fig:fig5}(a), the quantity of Ga2 vacancies on the \hkl(100), \hkl(001), and \hkl(-201) surfaces at 300 K exceeds that of Ga1 vacancies, aligning with the observations in the Ref.~\cite{he2024threshold}.
The number of Ga2 vacancies is lower than Ga1 vacancies at both 173 K and 500 K on the \hkl(100) surface.
On the \hkl(001) and \hkl(-201) surfaces, the quantities of Ga1 and Ga2 vacancies at 173 K are nearly identical.
In comparison, at 500 K, the results align with those observed at 300 K.
The \hkl(010) surface exhibits a unique characteristic wherein the quantity of Ga1 vacancies exceeds that of Ga2 vacancies at 173 K and 300 K.
On the (010) plane, the TDE of Ga1 and Ga2 vary at different sites, with the higher TDE exceeding 30 eV and the lower one falling below 20 eV~\cite{he2024threshold}.
The TDE of Ga2 on the \hkl(100) and \hkl(010) planes surpass that of Ga1.
This phenomenon is particularly pronounced for the \hkl(100) surface at low and high temperatures. In contrast, it tends to induce Ga1 with a smaller TDE by cascade collisions on the \hkl(010) plane.

For O atoms, the quantity of O vacancies at different lattice sites exhibits significant variation under varying conditions.
Furthermore, the distinctions between different O interstitials are markedly pronounced.
The highest abundance of O2 vacancies is observed on \hkl(100) and \hkl(010) surfaces for the lowest TDE, whereas the \hkl(001) and \hkl(-201) surfaces exhibit the lowest abundance because O2 has a higher TDE distribution than O1 and O3 according to Ref.~\cite{he2024threshold}.
Notably, fewer interstitial atoms are at the Ga2 site than at the Ga1 site.
This suggests that in the formation of antisite defects $\mathrm{Ga}_\mathrm{O}$, the Ga2 atom predominantly occupies the position held by the O1 atom.
Therefore, despite Ref.~\cite{he2024threshold} indicating TDE of O1 exceeding 60 eV, the O1 site predominantly hosts interstitials in \ce{Ga2O3} after irradiation.
Generally, Ga1 and O1 are the most common interstitial defects in the four surface orientations of $\beta$-\ce{Ga2O3}.
%The vacancy situation is more complicated.
The impact of temperature on the specific occupation of defects must also be evaluated with the surface orientation.

\subsection{Defects and carrier concentration variation} \label{sec:def}

\begin{figure}[ht!]
    \centering
    \includegraphics[width=8.6cm]{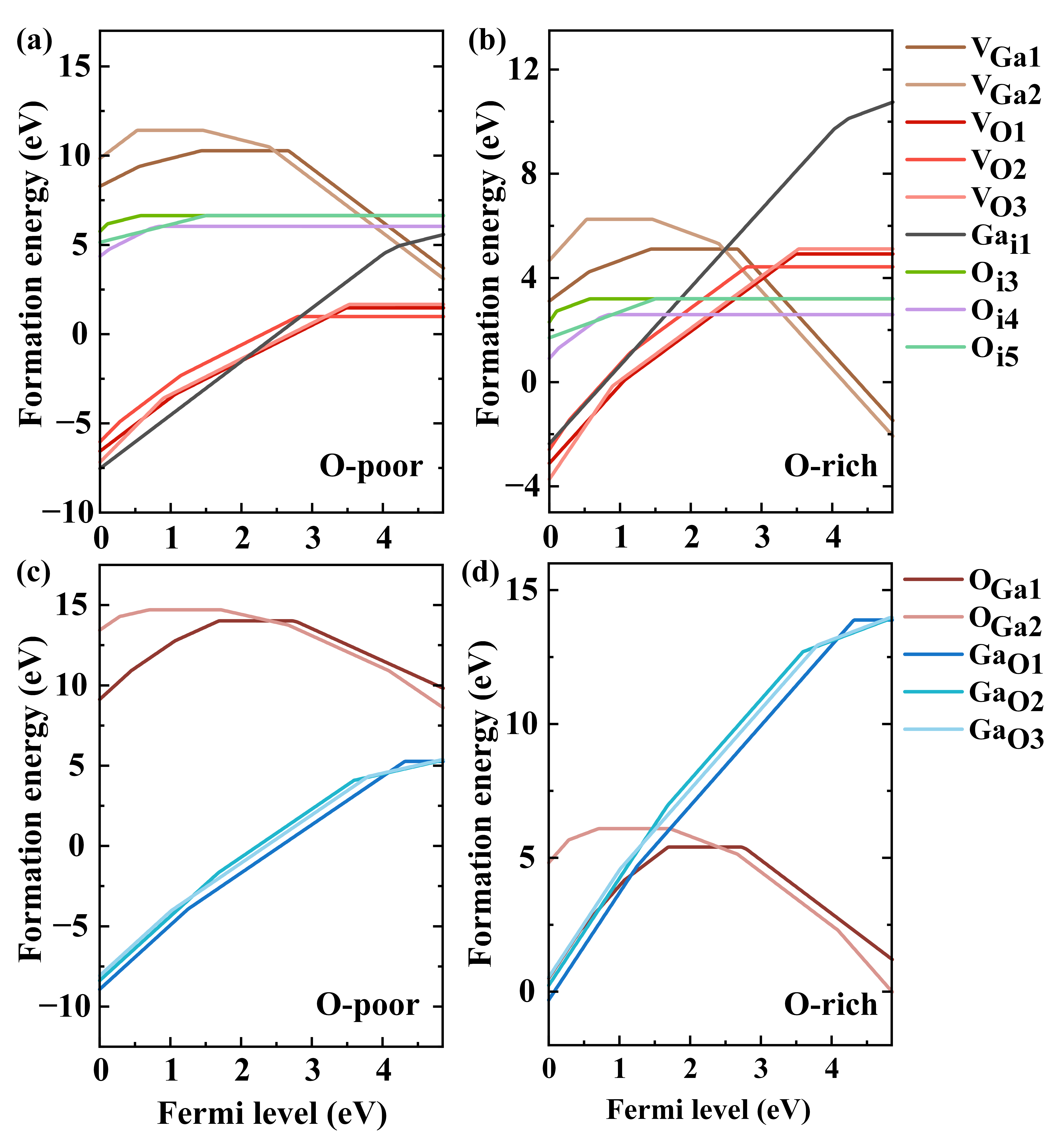}
    \caption{The formation energy of different defects in $\beta$-\ce{Ga2O3} at O-poor and O-rich limits.}
    \label{fig:fig6}
\end{figure}

For a deeper comprehension of the impact of radiation damage on $\beta$-\ce{Ga2O3}, the formation energy of point defects and antisite defects were calculated using DFT, as depicted in Fig.~\ref{fig:fig6}.
Only partial data are displayed here for Ga and O interstitials, and the complete data are in the supplementary material Figure~S8.
The transition level of Ga and O defects and antisite defects can be seen in the supplementary material Figures~S9 and S10, respectively. 
The $E_{f}$ for O vacancy is ranked as follows: $\text{O3} > \text{O1} > \text{O2}$. 
In Fig.~\ref{fig:fig6}(a), it is evident that O2 site tends to have more vacancies in most scenarios.
Given that the disparity in formation energy ($E_{f}$) between O1 and O3 is merely 0.5 eV, the variance in the defect quantities is not apparent.
In O-rich limits, the formation energy of O vacancy of O vacancy matches that found in $\mathrm{V}_\mathrm{Ga1}$ (Fig.~\ref{fig:fig6}(b)).
O interstitials possess the minimal $E_{f}$ and Ga interstitials exhibit the maximal.
This aligns with the findings in Fig.~\ref{fig:fig2},demonstrating O interstitial as the most abundant and Ga interstitial as the least prevalent.
In O-poor limits, the $E_{f}$ of Ga vacancies can exceed 10 eV, suggesting their influence can be neglected (Fig.~\ref{fig:fig6}(a)).
Conversely, O vacancies have the minimal $E_{f}$, indicating they are easily formed in this environment.
There is no notable variance in the $E_{f}$ between Ga and O interstitials.
It is worth noting that the overall $E_{f}$ of Ga vacancy is still more extensive than that of O, especially in O-poor conditions.
It makes sense that there should be more O vacancy defects than Ga vacancies.
However, the defect evolution after irradiation in Fig.~\ref{fig:fig2}(b) differs from the results.
As previously stated, this can be related to the formation of antisite defects in $\beta$-\ce{Ga2O3}.
In Fig.~\ref{fig:fig6}(c), the $\mathrm{Ga}_\mathrm{O}$ shows the lowest $E_{f}$, while $\mathrm{O}_\mathrm{Ga}$ exhibits the highest, peaking at 14 eV in an O-poor environment.
Hence, the $\mathrm{O}_\mathrm{Ga}$ defect can be ignored in such conditions.
O vacancy has a substantially lower defect formation energy than $\mathrm{Ga}_\mathrm{O}$, while Ga and O have similar interstitial $E_{f}$ with $\mathrm{Ga}_\mathrm{O}$.
The findings indicate a justifiable occurrence of $\mathrm{Ga}_\mathrm{O}$ defects.
In an O-rich environment, the defect formation energy of $\mathrm{Ga}_\mathrm{O}$ is much greater than that of $\mathrm{O}_\mathrm{Ga}$, as shown in Fig.~\ref{fig:fig6}(d).
$\mathrm{Ga}_\mathrm{O}$ defects are formed easily in O-poor conditions.
The defect formation energies of $\mathrm{O}_\mathrm{Ga1}$ and $\mathrm{O}_\mathrm{Ga2}$ are 5.4 eV and 6.1 eV, respectively. 
$\mathrm{O}_\mathrm{Ga1}$ is 0.3~eV higher than $\mathrm{V}_\mathrm{Ga1}$, and $\mathrm{O}_\mathrm{Ga2}$ is 0.16 eV lower than $\mathrm{V}_\mathrm{Ga2}$.
Results from $E_{f}$ are consistent with the ML-MD results, suggesting a greater likelihood of $\mathrm{Ga}_\mathrm{O}$ defect formation compared to $\mathrm{O}_\mathrm{Ga}$ defects.

A more in-depth examination of the electrical characteristics of various defect forms is justified.
As demonstrated in Figs.~\ref{fig:fig6}(a) and (b), O vacancy, Ga interstitial and antisite $\mathrm{Ga}_\mathrm{O}$ present the donor properties, consistent with the results of Refs.~\cite{HUANG2023_HSE_formationenergy, 2010varleyGa2O3}.
The transition energy level of defects is closer to the conduction band minimum (CBM), which makes it simpler to excite electron carriers.
In the supplementary material Figure~S9, the transition energy level of $\mathrm{V}_\mathrm{O2}$ about $+2/0$ is 1.32 eV below the CBM.
$\mathrm{V}_\mathrm{O1}$ and $\mathrm{V}_\mathrm{O3}$ are 1.36 eV and 2.06 eV below CBM, respectively.
The $4+$/$3+$ transition energy level of $\mathrm{V}_\mathrm{O2}$ is closest to VBM.
The transition energy level of 2+/1+ in $\mathrm{Ga}_\mathrm{i}$ and 3+/0 in $\mathrm{Ga}_\mathrm{O1}$ is 0.62 eV and 0.54 eV, respectively, below the CBM.
The primary compensation acceptor defects are Ga vacancies, with $\mathrm{V}_\mathrm{Ga2}$ being the nearest to the valence band maximum (VBM), featuring a transition energy level of 
 $3{+}/0$ positioned 0.52 eV above it in the supplementary material Figure~S9.
In O interstitials, the transition energy level of $+1/0$ in $\mathrm{O}_\mathrm{i3}$ and $\mathrm{O}_\mathrm{i4}$ is located at 0.57 eV and 0.84 eV above the VBM, respectively.
Other O interstitials are defects at a deeper level.
In O-poor limits, O vacancies have the lowest $E_{f}$, indicating that O vacancy defects are most likely to form in this environment.

\begin{figure*}[ht!]
    \centering
    \includegraphics[width=17cm]{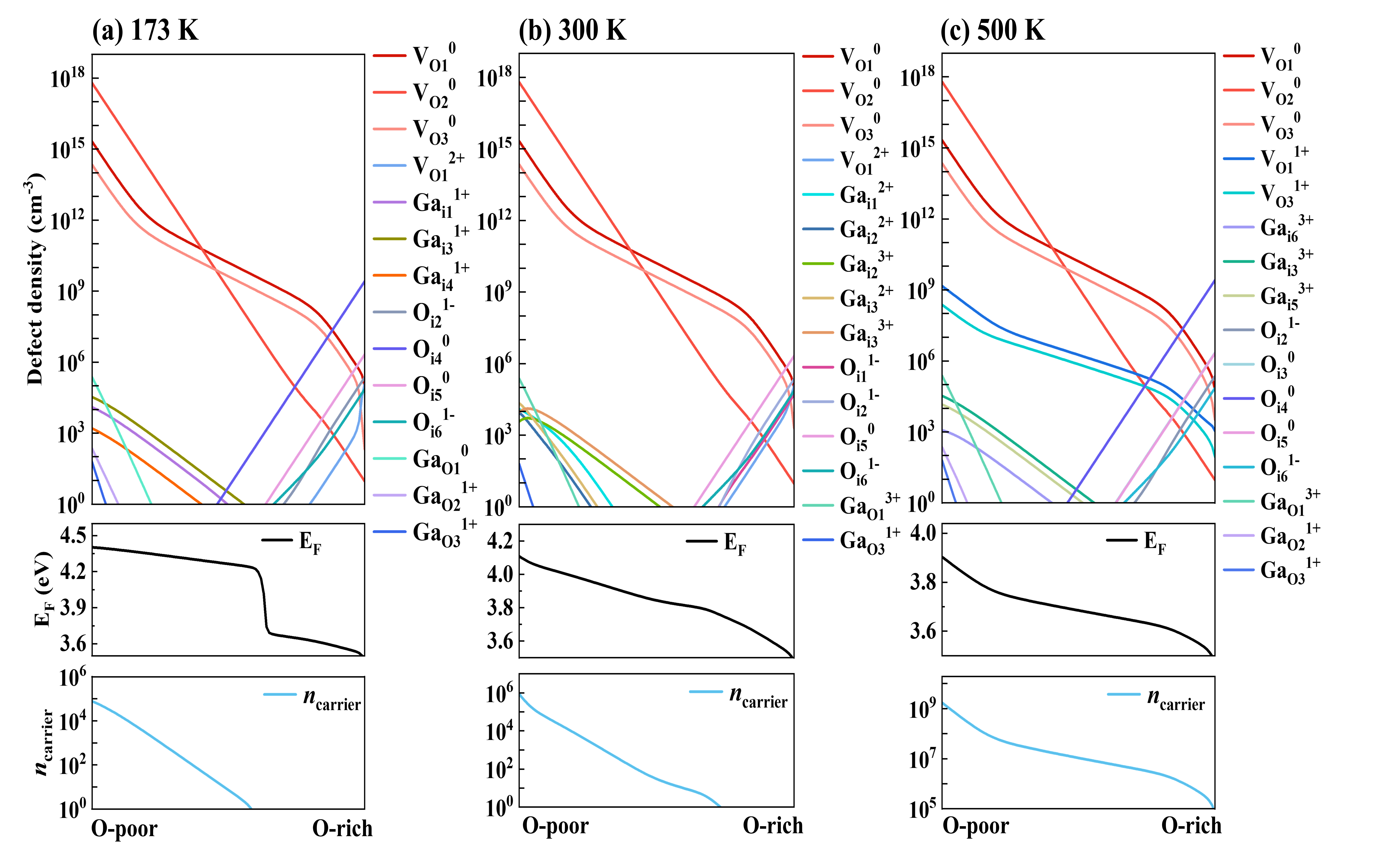}
    \caption{The defect density, Fermi level and carrier density at (a) 173 K, (b) 300 K and (c) 500~K in $\beta$-\ce{Ga2O3} films grown at the temperature of 1000 K as functions of the chemical potential, changing from O-poor to O-rich limits.}
    \label{fig:fig7}
\end{figure*}

Figure~\ref{fig:fig7} depicts the principal variations in defect concentration, Fermi level, and carrier concentration as chemical potential functions, assuming a growth temperature of 1000 K and operational temperatures of 173 K, 300 K, and 500 K, respectively.
The whole data is presented in the supplementary material Figures~S11-S13.
It is that O vacancy dominates at each temperature.
The concentration of $\mathrm{V}_\mathrm{O2}$ is the greatest, and $\mathrm{V}_\mathrm{O3}$ is the lowest, which can match the result of defect formation energy.
At 173 K, Ga interstitials and antisite defects $\mathrm{Ga}_\mathrm{O}$ emerge, and $\mathrm{Ga}_\mathrm{O1}^{0}$ is more concentrated than $\mathrm{Ga}_\mathrm{O2}^{1+}$ and $\mathrm{Ga}_\mathrm{O3}^{1+}$ in O-poor limits.
A reduction in neutral O vacancy concentration occurs as chemical potential nears an O-rich state, while there is an increase in $\mathrm{V}_\mathrm{O1}^{2+}$ and O interstitials.
Ga interstitial and $\mathrm{Ga}_\mathrm{O}$ with n-type conduction properties and O interstitials with deep acceptor features result in a low carrier concentration. %(Fig.~\ref{fig:fig6}), Fig.~S8 and Fig.~S9), resulting a low carrier concentration.
The Fermi level decreases as the chemical potential transitions from O-poor to O-rich.
This decline is characterized by a dip in the middle, attributable to low temperatures causing some carriers to become inactive.
More Ga and O interstitial defects are produced at 300 K, but no longer any $\mathrm{Ga}_\mathrm{O2}$.
Antisite defects are primarily composed of $\mathrm{Ga}_\mathrm{O1}^{3+}$ and $\mathrm{Ga}_\mathrm{O3}^{1+}$.
Since Ga interstitials show relatively shallow donor properties, the carrier concentration changes with chemical potential over a wider range than at 173 K.
Research on proton irradiation also demonstrated that antisite can impact the carrier lifetime of
$\beta$-\ce{Ga2O3}~\cite{2018ProtonII}.
At 500 K, there are more valence oxygen vacancies and a higher concentration of them.
Thus, the carrier concentration is higher than that at 173 K and 300 K.
$\mathrm{Ga}_\mathrm{O2}^{1+}$ appears with its concentration remaining approximately the same as at 173 K. 
An increase in the concentration of $\mathrm{O}_\mathrm{i4}^{0}$, coupled with a reduction in O vacancy, leads to a decrease in the carrier concentration.
The defect concentration of antisite defects $\mathrm{Ga}_\mathrm{O}$ is lower than the concentration of other point defects.
Specifically, the highest concentrations of O2 are found at 173 K and 300 K on the \hkl(100) surface, as well as 173 K and 500 K on the \hkl(010) surface in Fig.~\ref{fig:fig5}(a), which closely matches the findings in Fig.~\ref{fig:fig7}.
In the ML-MD simulation results in Fig.~\ref{fig:fig5}, there is a little more O3 than O2 at 300 K on the \hkl(010) surface.
The \hkl(001) surface exhibits the lowest concentration of O2 vacancy whereas the \hkl(-201) surface shows the highest concentration of O3 vacancy.
The different evolution of these defects will lead to variations in the electrical properties across different crystallographic planes.

\section{Conclusions}

In this work, the radiation damage mechanism of four surfaces \hkl(100), \hkl(010), \hkl(001), and \hkl(-201) of $\beta$-\ce{Ga2O3} at different temperatures of 173 K, 300 K, and 500 K was investigated by ML-MD simulation and DFT calculation.
The results indicate that following cascade collisions, the predominant defects of radiation damage on the four surfaces are Ga vacancies and O interstitials, with a propensity for the formation of antisite defect $\mathrm{Ga}_\mathrm{O}$.
Analysis of two Ga sites and three O sites across different surfaces reveals that the predominant intrinsic defects vary by surface, leading to distinct electronic properties.
In the specific lattice occupation of $\beta$-\ce{Ga2O3}, O2 site exhibits the highest concentration of oxygen atom.
The Ga1 and O1 sites have the highest concentration of interstitial atoms in Ga and O occupations, respectively.
$\mathrm{Ga}_\mathrm{O1}$ is the most likely antisite defect site.
Due to the significant channel effect, the \hkl(010) surface displays an extensive defect distribution and results in minimal defect generation.
The impact of temperature on the defect in radiation damage should be considered with surface orientation.
This finding is crucial for the rational design of $\beta$-\ce{Ga2O3}-based devices under extreme condition applications.

\subsection*{DECLARATION OF COMPETING INTEREST}
The authors declare that they have no known competing financial interests or personal relationships that could have appeared to influence the work reported in this paper.

\subsection*{ACKNOWLEDGMENTS}

The project was supported by the Major Program (JD) of Hubei Province (Grant No. 2023BAA009), the National Natural Science Foundation of China (Grant Nos. 52302046 and 62304097), the Knowledge Innovation Program of Wuhan-Shuguang (Grant No. 2023010201020262), the Natural Science Foundation of Jiangsu Province (Grant No. BK20230268), Guangdong Basic and Applied Basic Research Foundation (Grant No. 2023A1515012048), Shenzhen Fundamental Research Program (Grant Nos. JCYJ20230807093609019 and JCYJ20220530114615035), and the Open Fund of Hubei Key Laboratory of Electronic Manufacturing and Packaging Integration (Wuhan University) (Grant No. EMPI2024020). We gratefully acknowledge HZWTECH for providing computation facilities and thank Jie Li form HZWTECH for the help and discussions regarding this study. We also thank the Supercomputing Center of Wuhan University for their support of the calculation.

\bibliography{ref}

%apsrev4-2.bst 2019-01-14 (MD) hand-edited version of apsrev4-1.bst
%Control: key (0)
%Control: author (8) initials jnrlst
%Control: editor formatted (1) identically to author
%Control: production of article title (0) allowed
%Control: page (0) single
%Control: year (1) truncated
%Control: production of eprint (0) enabled
\begin{thebibliography}{53}%
\makeatletter
\providecommand \@ifxundefined [1]{%
 \@ifx{#1\undefined}
}%
\providecommand \@ifnum [1]{%
 \ifnum #1\expandafter \@firstoftwo
 \else \expandafter \@secondoftwo
 \fi
}%
\providecommand \@ifx [1]{%
 \ifx #1\expandafter \@firstoftwo
 \else \expandafter \@secondoftwo
 \fi
}%
\providecommand \natexlab [1]{#1}%
\providecommand \enquote  [1]{``#1''}%
\providecommand \bibnamefont  [1]{#1}%
\providecommand \bibfnamefont [1]{#1}%
\providecommand \citenamefont [1]{#1}%
\providecommand \href@noop [0]{\@secondoftwo}%
\providecommand \href [0]{\begingroup \@sanitize@url \@href}%
\providecommand \@href[1]{\@@startlink{#1}\@@href}%
\providecommand \@@href[1]{\endgroup#1\@@endlink}%
\providecommand \@sanitize@url [0]{\catcode `\\12\catcode `\$12\catcode `\&12\catcode `\#12\catcode `\^12\catcode `\_12\catcode `\%12\relax}%
\providecommand \@@startlink[1]{}%
\providecommand \@@endlink[0]{}%
\providecommand \url  [0]{\begingroup\@sanitize@url \@url }%
\providecommand \@url [1]{\endgroup\@href {#1}{\urlprefix }}%
\providecommand \urlprefix  [0]{URL }%
\providecommand \Eprint [0]{\href }%
\providecommand \doibase [0]{https://doi.org/}%
\providecommand \selectlanguage [0]{\@gobble}%
\providecommand \bibinfo  [0]{\@secondoftwo}%
\providecommand \bibfield  [0]{\@secondoftwo}%
\providecommand \translation [1]{[#1]}%
\providecommand \BibitemOpen [0]{}%
\providecommand \bibitemStop [0]{}%
\providecommand \bibitemNoStop [0]{.\EOS\space}%
\providecommand \EOS [0]{\spacefactor3000\relax}%
\providecommand \BibitemShut  [1]{\csname bibitem#1\endcsname}%
\let\auto@bib@innerbib\@empty
%</preamble>
\bibitem [{\citenamefont {Mastro}\ \emph {et~al.}(2017)\citenamefont {Mastro}, \citenamefont {Kuramata}, \citenamefont {Calkins}, \citenamefont {Kim}, \citenamefont {Ren},\ and\ \citenamefont {Pearton}}]{2017_perspective}%
  \BibitemOpen
  \bibfield  {author} {\bibinfo {author} {\bibfnamefont {M.~A.}\ \bibnamefont {Mastro}}, \bibinfo {author} {\bibfnamefont {A.}~\bibnamefont {Kuramata}}, \bibinfo {author} {\bibfnamefont {J.}~\bibnamefont {Calkins}}, \bibinfo {author} {\bibfnamefont {J.}~\bibnamefont {Kim}}, \bibinfo {author} {\bibfnamefont {F.}~\bibnamefont {Ren}},\ and\ \bibinfo {author} {\bibfnamefont {S.~J.}\ \bibnamefont {Pearton}},\ }\bibfield  {title} {\bibinfo {title} {Perspective—opportunities and future directions for \ce{Ga2O3}},\ }\href {https://doi.org/10.1149/2.0031707jss} {\bibfield  {journal} {\bibinfo  {journal} {ECS J. Solid State Sci. Technol.}\ }\textbf {\bibinfo {volume} {6}},\ \bibinfo {pages} {P356} (\bibinfo {year} {2017})}\BibitemShut {NoStop}%
\bibitem [{\citenamefont {Xie}\ \emph {et~al.}(2021)\citenamefont {Xie}, \citenamefont {Lu}, \citenamefont {Liang}, \citenamefont {Chen}, \citenamefont {Wang}, \citenamefont {Wu}, \citenamefont {Wu}, \citenamefont {Yang},\ and\ \citenamefont {Luo}}]{2021JMST}%
  \BibitemOpen
  \bibfield  {author} {\bibinfo {author} {\bibfnamefont {C.}~\bibnamefont {Xie}}, \bibinfo {author} {\bibfnamefont {X.}~\bibnamefont {Lu}}, \bibinfo {author} {\bibfnamefont {Y.}~\bibnamefont {Liang}}, \bibinfo {author} {\bibfnamefont {H.}~\bibnamefont {Chen}}, \bibinfo {author} {\bibfnamefont {L.}~\bibnamefont {Wang}}, \bibinfo {author} {\bibfnamefont {C.}~\bibnamefont {Wu}}, \bibinfo {author} {\bibfnamefont {D.}~\bibnamefont {Wu}}, \bibinfo {author} {\bibfnamefont {W.}~\bibnamefont {Yang}},\ and\ \bibinfo {author} {\bibfnamefont {L.}~\bibnamefont {Luo}},\ }\bibfield  {title} {\bibinfo {title} {Patterned growth of $\beta$-\ce{Ga2O3} thin films for solar-blind deep-ultraviolet photodetectors array and optical imaging application},\ }\href {https://doi.org/https://doi.org/10.1016/j.jmst.2020.09.015} {\bibfield  {journal} {\bibinfo  {journal} {J. Mater. Sci. Technol.}\ }\textbf {\bibinfo {volume} {72}},\ \bibinfo {pages} {189} (\bibinfo {year} {2021})}\BibitemShut {NoStop}%
\bibitem [{\citenamefont {Zhou}\ \emph {et~al.}(2024)\citenamefont {Zhou}, \citenamefont {Chen}, \citenamefont {Xu}, \citenamefont {Shao}, \citenamefont {Yang}, \citenamefont {Tian},\ and\ \citenamefont {Zhao}}]{2024RN}%
  \BibitemOpen
  \bibfield  {author} {\bibinfo {author} {\bibfnamefont {X.}~\bibnamefont {Zhou}}, \bibinfo {author} {\bibfnamefont {G.}~\bibnamefont {Chen}}, \bibinfo {author} {\bibfnamefont {L.}~\bibnamefont {Xu}}, \bibinfo {author} {\bibfnamefont {Z.}~\bibnamefont {Shao}}, \bibinfo {author} {\bibfnamefont {C.}~\bibnamefont {Yang}}, \bibinfo {author} {\bibfnamefont {Y.}~\bibnamefont {Tian}},\ and\ \bibinfo {author} {\bibfnamefont {Z.}~\bibnamefont {Zhao}},\ }\bibfield  {title} {\bibinfo {title} {A compact route for efficient production of high-purity $\beta$-\ce{Ga2O3} powder},\ }\href {https://doi.org/10.1007/s12598-024-02800-y} {\bibfield  {journal} {\bibinfo  {journal} {Rare Met.}\ }\textbf {\bibinfo {volume} {43}},\ \bibinfo {pages} {4573} (\bibinfo {year} {2024})}\BibitemShut {NoStop}%
\bibitem [{\citenamefont {Kim}\ \emph {et~al.}(2019)\citenamefont {Kim}, \citenamefont {Pearton}, \citenamefont {Fares}, \citenamefont {Yang}, \citenamefont {Ren}, \citenamefont {Kim},\ and\ \citenamefont {Polyakov}}]{2019RadiationEffectReview}%
  \BibitemOpen
  \bibfield  {author} {\bibinfo {author} {\bibfnamefont {J.}~\bibnamefont {Kim}}, \bibinfo {author} {\bibfnamefont {S.~J.}\ \bibnamefont {Pearton}}, \bibinfo {author} {\bibfnamefont {C.}~\bibnamefont {Fares}}, \bibinfo {author} {\bibfnamefont {J.}~\bibnamefont {Yang}}, \bibinfo {author} {\bibfnamefont {F.}~\bibnamefont {Ren}}, \bibinfo {author} {\bibfnamefont {S.}~\bibnamefont {Kim}},\ and\ \bibinfo {author} {\bibfnamefont {A.~Y.}\ \bibnamefont {Polyakov}},\ }\bibfield  {title} {\bibinfo {title} {Radiation damage effects in $\mathrm{Ga}_{2}\mathrm{O}_{3}$ materials and devices},\ }\href {https://doi.org/10.1039/C8TC04193H} {\bibfield  {journal} {\bibinfo  {journal} {J. Mater. Chem. C}\ }\textbf {\bibinfo {volume} {7}},\ \bibinfo {pages} {10} (\bibinfo {year} {2019})}\BibitemShut {NoStop}%
\bibitem [{\citenamefont {Petkov}\ \emph {et~al.}(2022)\citenamefont {Petkov}, \citenamefont {Cherns}, \citenamefont {Chen}, \citenamefont {Liu}, \citenamefont {Blevins}, \citenamefont {Gambin}, \citenamefont {Li}, \citenamefont {Liu},\ and\ \citenamefont {Kuball}}]{2022APLion_irradtaion}%
  \BibitemOpen
  \bibfield  {author} {\bibinfo {author} {\bibfnamefont {A.}~\bibnamefont {Petkov}}, \bibinfo {author} {\bibfnamefont {D.}~\bibnamefont {Cherns}}, \bibinfo {author} {\bibfnamefont {W.}~\bibnamefont {Chen}}, \bibinfo {author} {\bibfnamefont {J.}~\bibnamefont {Liu}}, \bibinfo {author} {\bibfnamefont {J.}~\bibnamefont {Blevins}}, \bibinfo {author} {\bibfnamefont {V.}~\bibnamefont {Gambin}}, \bibinfo {author} {\bibfnamefont {M.}~\bibnamefont {Li}}, \bibinfo {author} {\bibfnamefont {D.}~\bibnamefont {Liu}},\ and\ \bibinfo {author} {\bibfnamefont {M.}~\bibnamefont {Kuball}},\ }\bibfield  {title} {\bibinfo {title} {{Structural stability of $\beta$-\ce{Ga2O3} under ion irradiation}},\ }\href {https://doi.org/10.1063/5.0120089} {\bibfield  {journal} {\bibinfo  {journal} {Appl. Phys. Lett.}\ }\textbf {\bibinfo {volume} {121}},\ \bibinfo {pages} {171903} (\bibinfo {year} {2022})}\BibitemShut {NoStop}%
\bibitem [{\citenamefont {Azarov}\ \emph {et~al.}(2023)\citenamefont {Azarov}, \citenamefont {Fern{\'a}ndez}, \citenamefont {Zhao}, \citenamefont {Djurabekova}, \citenamefont {He}, \citenamefont {He}, \citenamefont {Prytz}, \citenamefont {Vines}, \citenamefont {Bektas}, \citenamefont {Chekhonin}, \citenamefont {Klingner}, \citenamefont {Hlawacek},\ and\ \citenamefont {Kuznetsov}}]{SFazarov2023universal}%
  \BibitemOpen
  \bibfield  {author} {\bibinfo {author} {\bibfnamefont {A.}~\bibnamefont {Azarov}}, \bibinfo {author} {\bibfnamefont {J.~G.}\ \bibnamefont {Fern{\'a}ndez}}, \bibinfo {author} {\bibfnamefont {J.}~\bibnamefont {Zhao}}, \bibinfo {author} {\bibfnamefont {F.}~\bibnamefont {Djurabekova}}, \bibinfo {author} {\bibfnamefont {H.}~\bibnamefont {He}}, \bibinfo {author} {\bibfnamefont {R.}~\bibnamefont {He}}, \bibinfo {author} {\bibfnamefont {{\O}.}~\bibnamefont {Prytz}}, \bibinfo {author} {\bibfnamefont {L.}~\bibnamefont {Vines}}, \bibinfo {author} {\bibfnamefont {U.}~\bibnamefont {Bektas}}, \bibinfo {author} {\bibfnamefont {P.}~\bibnamefont {Chekhonin}}, \bibinfo {author} {\bibfnamefont {N.}~\bibnamefont {Klingner}}, \bibinfo {author} {\bibfnamefont {G.}~\bibnamefont {Hlawacek}},\ and\ \bibinfo {author} {\bibfnamefont {A.}~\bibnamefont {Kuznetsov}},\ }\bibfield  {title} {\bibinfo {title} {Universal radiation tolerant semiconductor},\ }\href {https://doi.org/10.1038/s41467-023-40588-0} {\bibfield  {journal} {\bibinfo  {journal} {Nat. Commun.}\ }\textbf {\bibinfo {volume} {14}},\ \bibinfo {pages} {4855} (\bibinfo {year} {2023})}\BibitemShut {NoStop}%
\bibitem [{\citenamefont {De\'ak}\ \emph {et~al.}(2017)\citenamefont {De\'ak}, \citenamefont {Duy~Ho}, \citenamefont {Seemann}, \citenamefont {Aradi}, \citenamefont {Lorke},\ and\ \citenamefont {Frauenheim}}]{2017PeterDFTPhysRevB.95.075208}%
  \BibitemOpen
  \bibfield  {author} {\bibinfo {author} {\bibfnamefont {P.}~\bibnamefont {De\'ak}}, \bibinfo {author} {\bibfnamefont {Q.}~\bibnamefont {Duy~Ho}}, \bibinfo {author} {\bibfnamefont {F.}~\bibnamefont {Seemann}}, \bibinfo {author} {\bibfnamefont {B.}~\bibnamefont {Aradi}}, \bibinfo {author} {\bibfnamefont {M.}~\bibnamefont {Lorke}},\ and\ \bibinfo {author} {\bibfnamefont {T.}~\bibnamefont {Frauenheim}},\ }\bibfield  {title} {\bibinfo {title} {Choosing the correct hybrid for defect calculations: A case study on intrinsic carrier trapping in $\beta$-$\mathrm{Ga}_{2}\mathrm{O}_{3}$},\ }\href {https://doi.org/10.1103/PhysRevB.95.075208} {\bibfield  {journal} {\bibinfo  {journal} {Phys. Rev. B}\ }\textbf {\bibinfo {volume} {95}},\ \bibinfo {pages} {075208} (\bibinfo {year} {2017})}\BibitemShut {NoStop}%
\bibitem [{\citenamefont {Huang}\ \emph {et~al.}(2023{\natexlab{a}})\citenamefont {Huang}, \citenamefont {Xu}, \citenamefont {Yang}, \citenamefont {Yu}, \citenamefont {Wei}, \citenamefont {Ying}, \citenamefont {Liu}, \citenamefont {Jing}, \citenamefont {Li},\ and\ \citenamefont {Li}}]{2023IntrinsicDefects}%
  \BibitemOpen
  \bibfield  {author} {\bibinfo {author} {\bibfnamefont {Y.}~\bibnamefont {Huang}}, \bibinfo {author} {\bibfnamefont {X.}~\bibnamefont {Xu}}, \bibinfo {author} {\bibfnamefont {J.}~\bibnamefont {Yang}}, \bibinfo {author} {\bibfnamefont {X.}~\bibnamefont {Yu}}, \bibinfo {author} {\bibfnamefont {Y.}~\bibnamefont {Wei}}, \bibinfo {author} {\bibfnamefont {T.}~\bibnamefont {Ying}}, \bibinfo {author} {\bibfnamefont {Z.}~\bibnamefont {Liu}}, \bibinfo {author} {\bibfnamefont {Y.}~\bibnamefont {Jing}}, \bibinfo {author} {\bibfnamefont {W.}~\bibnamefont {Li}},\ and\ \bibinfo {author} {\bibfnamefont {X.}~\bibnamefont {Li}},\ }\bibfield  {title} {\bibinfo {title} {Library of intrinsic defects in $\beta$-\ce{Ga2O3}: First-principles studies},\ }\href {https://doi.org/https://doi.org/10.1016/j.mtcomm.2023.105898} {\bibfield  {journal} {\bibinfo  {journal} {Mater. Today Commun.}\ }\textbf {\bibinfo {volume} {35}},\ \bibinfo {pages} {105898} (\bibinfo {year} {2023}{\natexlab{a}})}\BibitemShut {NoStop}%
\bibitem [{\citenamefont {Varley}\ \emph {et~al.}(2010)\citenamefont {Varley}, \citenamefont {Weber}, \citenamefont {Janotti},\ and\ \citenamefont {Van~de Walle}}]{2010varleyGa2O3}%
  \BibitemOpen
  \bibfield  {author} {\bibinfo {author} {\bibfnamefont {J.~B.}\ \bibnamefont {Varley}}, \bibinfo {author} {\bibfnamefont {J.~R.}\ \bibnamefont {Weber}}, \bibinfo {author} {\bibfnamefont {A.}~\bibnamefont {Janotti}},\ and\ \bibinfo {author} {\bibfnamefont {C.~G.}\ \bibnamefont {Van~de Walle}},\ }\bibfield  {title} {\bibinfo {title} {{Oxygen vacancies and donor impurities in $\beta$-$\mathrm{Ga}_{2}\mathrm{O}_{3}$}},\ }\href {https://doi.org/10.1063/1.3499306} {\bibfield  {journal} {\bibinfo  {journal} {Appl. Phys. Lett.}\ }\textbf {\bibinfo {volume} {97}},\ \bibinfo {pages} {142106} (\bibinfo {year} {2010})}\BibitemShut {NoStop}%
\bibitem [{\citenamefont {Dong}\ \emph {et~al.}(2017)\citenamefont {Dong}, \citenamefont {Jia}, \citenamefont {Xin}, \citenamefont {Peng},\ and\ \citenamefont {Zhang}}]{2017Luminescence}%
  \BibitemOpen
  \bibfield  {author} {\bibinfo {author} {\bibfnamefont {L.}~\bibnamefont {Dong}}, \bibinfo {author} {\bibfnamefont {R.}~\bibnamefont {Jia}}, \bibinfo {author} {\bibfnamefont {B.}~\bibnamefont {Xin}}, \bibinfo {author} {\bibfnamefont {B.}~\bibnamefont {Peng}},\ and\ \bibinfo {author} {\bibfnamefont {Y.}~\bibnamefont {Zhang}},\ }\bibfield  {title} {\bibinfo {title} {Effects of oxygen vacancies on the structural and optical properties of $\beta$-\ce{Ga2O3}},\ }\href {https://doi.org/10.1038/srep40160} {\bibfield  {journal} {\bibinfo  {journal} {Sci. Rep.}\ }\textbf {\bibinfo {volume} {7}},\ \bibinfo {pages} {40160} (\bibinfo {year} {2017})}\BibitemShut {NoStop}%
\bibitem [{\citenamefont {{Cojocaru}}(1974)}]{1974RadEf}%
  \BibitemOpen
  \bibfield  {author} {\bibinfo {author} {\bibfnamefont {L.~N.}\ \bibnamefont {{Cojocaru}}},\ }\bibfield  {title} {\bibinfo {title} {{Defect-annealing in neutron-damaged $\beta$-\ce{Ga2O3}}},\ }\href {https://doi.org/10.1080/00337577408241456} {\bibfield  {journal} {\bibinfo  {journal} {Radiation Effects}\ }\textbf {\bibinfo {volume} {21}},\ \bibinfo {pages} {157} (\bibinfo {year} {1974})}\BibitemShut {NoStop}%
\bibitem [{\citenamefont {Ingebrigtsen}\ \emph {et~al.}(2018)\citenamefont {Ingebrigtsen}, \citenamefont {Kuznetsov}, \citenamefont {Svensson}, \citenamefont {Alfieri}, \citenamefont {Mihaila}, \citenamefont {Badstübner}, \citenamefont {Perron}, \citenamefont {Vines},\ and\ \citenamefont {Varley}}]{2019proton}%
  \BibitemOpen
  \bibfield  {author} {\bibinfo {author} {\bibfnamefont {M.~E.}\ \bibnamefont {Ingebrigtsen}}, \bibinfo {author} {\bibfnamefont {A.~Y.}\ \bibnamefont {Kuznetsov}}, \bibinfo {author} {\bibfnamefont {B.~G.}\ \bibnamefont {Svensson}}, \bibinfo {author} {\bibfnamefont {G.}~\bibnamefont {Alfieri}}, \bibinfo {author} {\bibfnamefont {A.}~\bibnamefont {Mihaila}}, \bibinfo {author} {\bibfnamefont {U.}~\bibnamefont {Badstübner}}, \bibinfo {author} {\bibfnamefont {A.}~\bibnamefont {Perron}}, \bibinfo {author} {\bibfnamefont {L.}~\bibnamefont {Vines}},\ and\ \bibinfo {author} {\bibfnamefont {J.~B.}\ \bibnamefont {Varley}},\ }\bibfield  {title} {\bibinfo {title} {{Impact of proton irradiation on conductivity and deep level defects in $\beta$-\ce{Ga2O3}}},\ }\href {https://doi.org/10.1063/1.5054826} {\bibfield  {journal} {\bibinfo  {journal} {APL Mater.}\ }\textbf {\bibinfo {volume} {7}},\ \bibinfo {pages} {022510} (\bibinfo {year} {2018})}\BibitemShut {NoStop}%
\bibitem [{\citenamefont {Lee}\ \emph {et~al.}(2018)\citenamefont {Lee}, \citenamefont {Flitsiyan}, \citenamefont {Chernyak}, \citenamefont {Yang}, \citenamefont {Ren}, \citenamefont {Pearton}, \citenamefont {Meyler},\ and\ \citenamefont {Salzman}}]{2018Lee}%
  \BibitemOpen
  \bibfield  {author} {\bibinfo {author} {\bibfnamefont {J.}~\bibnamefont {Lee}}, \bibinfo {author} {\bibfnamefont {E.}~\bibnamefont {Flitsiyan}}, \bibinfo {author} {\bibfnamefont {L.}~\bibnamefont {Chernyak}}, \bibinfo {author} {\bibfnamefont {J.}~\bibnamefont {Yang}}, \bibinfo {author} {\bibfnamefont {F.}~\bibnamefont {Ren}}, \bibinfo {author} {\bibfnamefont {S.~J.}\ \bibnamefont {Pearton}}, \bibinfo {author} {\bibfnamefont {B.}~\bibnamefont {Meyler}},\ and\ \bibinfo {author} {\bibfnamefont {Y.~J.}\ \bibnamefont {Salzman}},\ }\bibfield  {title} {\bibinfo {title} {{Effect of 1.5 MeV electron irradiation on $\beta$-\ce{Ga2O3} carrier lifetime and diffusion length}},\ }\href {https://doi.org/10.1063/1.5011971} {\bibfield  {journal} {\bibinfo  {journal} {Appl. Phys. Lett.}\ }\textbf {\bibinfo {volume} {112}},\ \bibinfo {pages} {082104} (\bibinfo {year} {2018})}\BibitemShut {NoStop}%
\bibitem [{\citenamefont {Zhang}\ \emph {et~al.}(2010)\citenamefont {Zhang}, \citenamefont {Fu}, \citenamefont {Misra},\ and\ \citenamefont {Demkowicz}}]{2010Interface-defect}%
  \BibitemOpen
  \bibfield  {author} {\bibinfo {author} {\bibfnamefont {X.}~\bibnamefont {Zhang}}, \bibinfo {author} {\bibfnamefont {E.~G.}\ \bibnamefont {Fu}}, \bibinfo {author} {\bibfnamefont {A.}~\bibnamefont {Misra}},\ and\ \bibinfo {author} {\bibfnamefont {M.~J.}\ \bibnamefont {Demkowicz}},\ }\bibfield  {title} {\bibinfo {title} {Interface-enabled defect reduction in he ion irradiated metallic multilayers},\ }\href {https://doi.org/10.1007/s11837-010-0185-5} {\bibfield  {journal} {\bibinfo  {journal} {JOM}\ }\textbf {\bibinfo {volume} {62}},\ \bibinfo {pages} {75} (\bibinfo {year} {2010})}\BibitemShut {NoStop}%
\bibitem [{\citenamefont {Yang}\ \emph {et~al.}(2021)\citenamefont {Yang}, \citenamefont {Tang}, \citenamefont {Zhang}, \citenamefont {Ma}, \citenamefont {Liu}, \citenamefont {Li}, \citenamefont {Zhang}, \citenamefont {Li}, \citenamefont {Liu}, \citenamefont {Fan},\ and\ \citenamefont {Namakian}}]{2021YANGinterface-defect}%
  \BibitemOpen
  \bibfield  {author} {\bibinfo {author} {\bibfnamefont {K.}~\bibnamefont {Yang}}, \bibinfo {author} {\bibfnamefont {P.}~\bibnamefont {Tang}}, \bibinfo {author} {\bibfnamefont {Q.}~\bibnamefont {Zhang}}, \bibinfo {author} {\bibfnamefont {H.}~\bibnamefont {Ma}}, \bibinfo {author} {\bibfnamefont {E.}~\bibnamefont {Liu}}, \bibinfo {author} {\bibfnamefont {M.}~\bibnamefont {Li}}, \bibinfo {author} {\bibfnamefont {X.}~\bibnamefont {Zhang}}, \bibinfo {author} {\bibfnamefont {J.}~\bibnamefont {Li}}, \bibinfo {author} {\bibfnamefont {Y.}~\bibnamefont {Liu}}, \bibinfo {author} {\bibfnamefont {T.}~\bibnamefont {Fan}},\ and\ \bibinfo {author} {\bibfnamefont {R.}~\bibnamefont {Namakian}},\ }\bibfield  {title} {\bibinfo {title} {Enhanced defect annihilation capability of the graphene/copper interface: An in situ study},\ }\href {https://doi.org/https://doi.org/10.1016/j.scriptamat.2021.114001} {\bibfield  {journal} {\bibinfo  {journal} {Scr. Mater.}\ }\textbf {\bibinfo {volume} {203}},\ \bibinfo {pages} {114001} (\bibinfo {year} {2021})}\BibitemShut {NoStop}%
\bibitem [{\citenamefont {Wei}\ \emph {et~al.}(2018)\citenamefont {Wei}, \citenamefont {Ren}, \citenamefont {Qin}, \citenamefont {Hu}, \citenamefont {Deng},\ and\ \citenamefont {Jiang}}]{Weiguo_W2018}%
  \BibitemOpen
  \bibfield  {author} {\bibinfo {author} {\bibfnamefont {G.}~\bibnamefont {Wei}}, \bibinfo {author} {\bibfnamefont {F.}~\bibnamefont {Ren}}, \bibinfo {author} {\bibfnamefont {W.}~\bibnamefont {Qin}}, \bibinfo {author} {\bibfnamefont {W.}~\bibnamefont {Hu}}, \bibinfo {author} {\bibfnamefont {H.}~\bibnamefont {Deng}},\ and\ \bibinfo {author} {\bibfnamefont {C.}~\bibnamefont {Jiang}},\ }\bibfield  {title} {\bibinfo {title} {Evolution of helium bubbles below different tungsten surfaces under neutron irradiation and non-irradiation conditions},\ }\href {https://doi.org/https://doi.org/10.1016/j.commatsci.2018.02.050} {\bibfield  {journal} {\bibinfo  {journal} {Comput. Mater. Sci.}\ }\textbf {\bibinfo {volume} {148}},\ \bibinfo {pages} {242} (\bibinfo {year} {2018})}\BibitemShut {NoStop}%
\bibitem [{\citenamefont {Wei}\ \emph {et~al.}(2019)\citenamefont {Wei}, \citenamefont {Ren}, \citenamefont {Fang}, \citenamefont {Hu}, \citenamefont {Gao}, \citenamefont {Qin}, \citenamefont {Cheng}, \citenamefont {Wang}, \citenamefont {Jiang},\ and\ \citenamefont {Deng}}]{Weiguo_2019}%
  \BibitemOpen
  \bibfield  {author} {\bibinfo {author} {\bibfnamefont {G.}~\bibnamefont {Wei}}, \bibinfo {author} {\bibfnamefont {F.}~\bibnamefont {Ren}}, \bibinfo {author} {\bibfnamefont {J.}~\bibnamefont {Fang}}, \bibinfo {author} {\bibfnamefont {W.}~\bibnamefont {Hu}}, \bibinfo {author} {\bibfnamefont {F.}~\bibnamefont {Gao}}, \bibinfo {author} {\bibfnamefont {W.}~\bibnamefont {Qin}}, \bibinfo {author} {\bibfnamefont {T.}~\bibnamefont {Cheng}}, \bibinfo {author} {\bibfnamefont {Y.}~\bibnamefont {Wang}}, \bibinfo {author} {\bibfnamefont {C.}~\bibnamefont {Jiang}},\ and\ \bibinfo {author} {\bibfnamefont {H.}~\bibnamefont {Deng}},\ }\bibfield  {title} {\bibinfo {title} {Understanding the release of helium atoms from nanochannel tungsten: a molecular dynamics simulation},\ }\href {https://doi.org/10.1088/1741-4326/ab14c7} {\bibfield  {journal} {\bibinfo  {journal} {Nucl. Fusion}\ }\textbf {\bibinfo {volume} {59}},\ \bibinfo {pages} {076020} (\bibinfo {year} {2019})}\BibitemShut {NoStop}%
\bibitem [{\citenamefont {Liu}\ \emph {et~al.}(2022)\citenamefont {Liu}, \citenamefont {Shao}, \citenamefont {Lyu}, \citenamefont {Lai},\ and\ \citenamefont {Shen}}]{Liu_2022}%
  \BibitemOpen
  \bibfield  {author} {\bibinfo {author} {\bibfnamefont {T.}~\bibnamefont {Liu}}, \bibinfo {author} {\bibfnamefont {T.}~\bibnamefont {Shao}}, \bibinfo {author} {\bibfnamefont {F.}~\bibnamefont {Lyu}}, \bibinfo {author} {\bibfnamefont {X.}~\bibnamefont {Lai}},\ and\ \bibinfo {author} {\bibfnamefont {A.~H.}\ \bibnamefont {Shen}},\ }\bibfield  {title} {\bibinfo {title} {Molecular dynamics simulations to assess the radiation resistance of different crystal orientations of diamond under neutron irradiation},\ }\href {https://doi.org/10.1088/1361-651X/ac4c98} {\bibfield  {journal} {\bibinfo  {journal} {Modell. Simul. Mater. Sci. Eng.}\ }\textbf {\bibinfo {volume} {30}},\ \bibinfo {pages} {035005} (\bibinfo {year} {2022})}\BibitemShut {NoStop}%
\bibitem [{\citenamefont {Charnvanichborikarn}\ \emph {et~al.}(2012)\citenamefont {Charnvanichborikarn}, \citenamefont {Myers}, \citenamefont {Shao},\ and\ \citenamefont {Kucheyev}}]{2012_GaNSurfaceDamage}%
  \BibitemOpen
  \bibfield  {author} {\bibinfo {author} {\bibfnamefont {S.}~\bibnamefont {Charnvanichborikarn}}, \bibinfo {author} {\bibfnamefont {M.}~\bibnamefont {Myers}}, \bibinfo {author} {\bibfnamefont {L.}~\bibnamefont {Shao}},\ and\ \bibinfo {author} {\bibfnamefont {S.}~\bibnamefont {Kucheyev}},\ }\bibfield  {title} {\bibinfo {title} {Interface-mediated suppression of radiation damage in \ce{GaN}},\ }\href {https://doi.org/https://doi.org/10.1016/j.scriptamat.2012.04.020} {\bibfield  {journal} {\bibinfo  {journal} {Scr. Mater.}\ }\textbf {\bibinfo {volume} {67}},\ \bibinfo {pages} {205} (\bibinfo {year} {2012})}\BibitemShut {NoStop}%
\bibitem [{\citenamefont {Huang}\ \emph {et~al.}(2024)\citenamefont {Huang}, \citenamefont {Xu}, \citenamefont {Yang}, \citenamefont {Yu}, \citenamefont {Wei}, \citenamefont {Ying}, \citenamefont {Liu}, \citenamefont {Jing}, \citenamefont {Li},\ and\ \citenamefont {Li}}]{2024SchottkyBarrierIrradiation}%
  \BibitemOpen
  \bibfield  {author} {\bibinfo {author} {\bibfnamefont {Y.}~\bibnamefont {Huang}}, \bibinfo {author} {\bibfnamefont {X.}~\bibnamefont {Xu}}, \bibinfo {author} {\bibfnamefont {J.}~\bibnamefont {Yang}}, \bibinfo {author} {\bibfnamefont {X.}~\bibnamefont {Yu}}, \bibinfo {author} {\bibfnamefont {Y.}~\bibnamefont {Wei}}, \bibinfo {author} {\bibfnamefont {T.}~\bibnamefont {Ying}}, \bibinfo {author} {\bibfnamefont {Z.}~\bibnamefont {Liu}}, \bibinfo {author} {\bibfnamefont {Y.}~\bibnamefont {Jing}}, \bibinfo {author} {\bibfnamefont {W.}~\bibnamefont {Li}},\ and\ \bibinfo {author} {\bibfnamefont {X.}~\bibnamefont {Li}},\ }\bibfield  {title} {\bibinfo {title} {Defect identification in $\beta$-\ce{Ga2O3} schottky barrier diodes with electron radiation and annealing regulating},\ }\href {https://doi.org/10.1109/TNS.2024.3383441} {\bibfield  {journal} {\bibinfo  {journal} {IEEE Trans. Nucl. Sci}\ }\textbf {\bibinfo {volume} {71}},\ \bibinfo {pages} {1178} (\bibinfo {year} {2024})}\BibitemShut {NoStop}%
\bibitem [{\citenamefont {Blevins}\ \emph {et~al.}(2019)\citenamefont {Blevins}, \citenamefont {Stevens}, \citenamefont {Lindsey}, \citenamefont {Foundos},\ and\ \citenamefont {Sande}}]{2019Large(100)}%
  \BibitemOpen
  \bibfield  {author} {\bibinfo {author} {\bibfnamefont {J.~D.}\ \bibnamefont {Blevins}}, \bibinfo {author} {\bibfnamefont {K.}~\bibnamefont {Stevens}}, \bibinfo {author} {\bibfnamefont {A.}~\bibnamefont {Lindsey}}, \bibinfo {author} {\bibfnamefont {G.}~\bibnamefont {Foundos}},\ and\ \bibinfo {author} {\bibfnamefont {L.}~\bibnamefont {Sande}},\ }\bibfield  {title} {\bibinfo {title} {Development of large diameter semi-insulating gallium oxide ($\mathrm{Ga}_{2}\mathrm{O}_{3}$) substrates},\ }\href {https://doi.org/10.1109/TSM.2019.2944526} {\bibfield  {journal} {\bibinfo  {journal} {IEEE Trans. Semicond. Manuf.}\ }\textbf {\bibinfo {volume} {32}},\ \bibinfo {pages} {466} (\bibinfo {year} {2019})}\BibitemShut {NoStop}%
\bibitem [{\citenamefont {{Bermudez}}(2006)}]{2006Low-indexSufaces}%
  \BibitemOpen
  \bibfield  {author} {\bibinfo {author} {\bibfnamefont {V.~M.}\ \bibnamefont {{Bermudez}}},\ }\bibfield  {title} {\bibinfo {title} {{The structure of low-index surfaces of $\beta$-\ce{Ga2O3}}},\ }\href {https://doi.org/10.1016/j.chemphys.2005.08.051} {\bibfield  {journal} {\bibinfo  {journal} {Chem. Phys.}\ }\textbf {\bibinfo {volume} {323}},\ \bibinfo {pages} {193} (\bibinfo {year} {2006})}\BibitemShut {NoStop}%
\bibitem [{\citenamefont {Mu}\ \emph {et~al.}(2020)\citenamefont {Mu}, \citenamefont {Wang}, \citenamefont {Peelaers},\ and\ \citenamefont {Van~de Walle}}]{MuSai2020(100)}%
  \BibitemOpen
  \bibfield  {author} {\bibinfo {author} {\bibfnamefont {S.}~\bibnamefont {Mu}}, \bibinfo {author} {\bibfnamefont {M.}~\bibnamefont {Wang}}, \bibinfo {author} {\bibfnamefont {H.}~\bibnamefont {Peelaers}},\ and\ \bibinfo {author} {\bibfnamefont {C.~G.}\ \bibnamefont {Van~de Walle}},\ }\bibfield  {title} {\bibinfo {title} {{First-principles surface energies for monoclinic $\mathrm{Ga}_{2}\mathrm{O}_{3}$ and \ce{Al2O3} and consequences for cracking of $(\mathrm{Al}_{x}\mathrm{Ga}_{1-x})_{2}\mathrm{O}_{3}$}},\ }\href {https://doi.org/10.1063/5.0019915} {\bibfield  {journal} {\bibinfo  {journal} {APL Mater.}\ }\textbf {\bibinfo {volume} {8}},\ \bibinfo {pages} {091105} (\bibinfo {year} {2020})}\BibitemShut {NoStop}%
\bibitem [{\citenamefont {Sasaki}\ \emph {et~al.}(2012)\citenamefont {Sasaki}, \citenamefont {Kuramata}, \citenamefont {Masui}, \citenamefont {Víllora}, \citenamefont {Shimamura},\ and\ \citenamefont {Yamakoshi}}]{Sasaki_2012(010)surface}%
  \BibitemOpen
  \bibfield  {author} {\bibinfo {author} {\bibfnamefont {K.}~\bibnamefont {Sasaki}}, \bibinfo {author} {\bibfnamefont {A.}~\bibnamefont {Kuramata}}, \bibinfo {author} {\bibfnamefont {T.}~\bibnamefont {Masui}}, \bibinfo {author} {\bibfnamefont {E.~G.}\ \bibnamefont {Víllora}}, \bibinfo {author} {\bibfnamefont {K.}~\bibnamefont {Shimamura}},\ and\ \bibinfo {author} {\bibfnamefont {S.}~\bibnamefont {Yamakoshi}},\ }\bibfield  {title} {\bibinfo {title} {Device-quality $\beta$-$\mathrm{Ga}_{2}\mathrm{O}_{3}$ epitaxial films fabricated by ozone molecular beam epitaxy},\ }\href {https://doi.org/10.1143/APEX.5.035502} {\bibfield  {journal} {\bibinfo  {journal} {Appl. Phys. Express}\ }\textbf {\bibinfo {volume} {5}},\ \bibinfo {pages} {035502} (\bibinfo {year} {2012})}\BibitemShut {NoStop}%
\bibitem [{\citenamefont {Mastro}\ \emph {et~al.}(2020)\citenamefont {Mastro}, \citenamefont {Eddy}, \citenamefont {Tadjer}, \citenamefont {Hite}, \citenamefont {Kim},\ and\ \citenamefont {Pearton}}]{2020_010surface}%
  \BibitemOpen
  \bibfield  {author} {\bibinfo {author} {\bibfnamefont {M.~A.}\ \bibnamefont {Mastro}}, \bibinfo {author} {\bibfnamefont {J.}~\bibnamefont {Eddy}, \bibfnamefont {Charles~R.}}, \bibinfo {author} {\bibfnamefont {M.~J.}\ \bibnamefont {Tadjer}}, \bibinfo {author} {\bibfnamefont {J.~K.}\ \bibnamefont {Hite}}, \bibinfo {author} {\bibfnamefont {J.}~\bibnamefont {Kim}},\ and\ \bibinfo {author} {\bibfnamefont {S.~J.}\ \bibnamefont {Pearton}},\ }\bibfield  {title} {\bibinfo {title} {{Assessment of the \hkl(010) $\beta$-\ce{Ga2O3} surface and substrate specification}},\ }\href {https://doi.org/10.1116/6.0000725} {\bibfield  {journal} {\bibinfo  {journal} {J. Vac. Sci. Technol. A}\ }\textbf {\bibinfo {volume} {39}},\ \bibinfo {pages} {013408} (\bibinfo {year} {2020})}\BibitemShut {NoStop}%
\bibitem [{\citenamefont {Tadjer}\ \emph {et~al.}(2020)\citenamefont {Tadjer}, \citenamefont {Alema}, \citenamefont {Osinsky}, \citenamefont {Mastro}, \citenamefont {Nepal}, \citenamefont {Woodward}, \citenamefont {Myers-Ward}, \citenamefont {Glaser}, \citenamefont {Freitas}, \citenamefont {Jacobs}, \citenamefont {Gallagher}, \citenamefont {Mock}, \citenamefont {Pennachio}, \citenamefont {Hajzus}, \citenamefont {Ebrish}, \citenamefont {Anderson}, \citenamefont {Hobart}, \citenamefont {Hite},\ and\ \citenamefont {Jr.}}]{Tadjer_2021}%
  \BibitemOpen
  \bibfield  {author} {\bibinfo {author} {\bibfnamefont {M.~J.}\ \bibnamefont {Tadjer}}, \bibinfo {author} {\bibfnamefont {F.}~\bibnamefont {Alema}}, \bibinfo {author} {\bibfnamefont {A.}~\bibnamefont {Osinsky}}, \bibinfo {author} {\bibfnamefont {M.~A.}\ \bibnamefont {Mastro}}, \bibinfo {author} {\bibfnamefont {N.}~\bibnamefont {Nepal}}, \bibinfo {author} {\bibfnamefont {J.~M.}\ \bibnamefont {Woodward}}, \bibinfo {author} {\bibfnamefont {R.~L.}\ \bibnamefont {Myers-Ward}}, \bibinfo {author} {\bibfnamefont {E.~R.}\ \bibnamefont {Glaser}}, \bibinfo {author} {\bibfnamefont {J.~A.}\ \bibnamefont {Freitas}}, \bibinfo {author} {\bibfnamefont {A.~G.}\ \bibnamefont {Jacobs}}, \bibinfo {author} {\bibfnamefont {J.~C.}\ \bibnamefont {Gallagher}}, \bibinfo {author} {\bibfnamefont {A.~L.}\ \bibnamefont {Mock}}, \bibinfo {author} {\bibfnamefont {D.~J.}\ \bibnamefont {Pennachio}}, \bibinfo {author} {\bibfnamefont {J.}~\bibnamefont {Hajzus}}, \bibinfo {author} {\bibfnamefont {M.}~\bibnamefont {Ebrish}}, \bibinfo {author} {\bibfnamefont {T.~J.}\ \bibnamefont {Anderson}}, \bibinfo {author} {\bibfnamefont {K.~D.}\ \bibnamefont {Hobart}}, \bibinfo {author} {\bibfnamefont {J.~K.}\ \bibnamefont {Hite}},\ and\ \bibinfo {author} {\bibfnamefont {C.~R.~E.}\ \bibnamefont {Jr.}},\ }\bibfield  {title} {\bibinfo {title} {Characterization of $\beta$-\ce{Ga2O3} homoepitaxial films and mosfets grown by mocvd at high growth rates},\ }\href {https://doi.org/10.1088/1361-6463/abbc96} {\bibfield  {journal} {\bibinfo  {journal} {J. Phys. D: Appl. Phys.}\ }\textbf {\bibinfo {volume} {54}},\ \bibinfo {pages} {034005} (\bibinfo {year} {2020})}\BibitemShut {NoStop}%
\bibitem [{\citenamefont {Wang}\ \emph {et~al.}(2023)\citenamefont {Wang}, \citenamefont {Mu}, \citenamefont {Speck},\ and\ \citenamefont {Van~de Walle}}]{WANG-dft2023}%
  \BibitemOpen
  \bibfield  {author} {\bibinfo {author} {\bibfnamefont {M.}~\bibnamefont {Wang}}, \bibinfo {author} {\bibfnamefont {S.}~\bibnamefont {Mu}}, \bibinfo {author} {\bibfnamefont {J.~S.}\ \bibnamefont {Speck}},\ and\ \bibinfo {author} {\bibfnamefont {C.~G.}\ \bibnamefont {Van~de Walle}},\ }\bibfield  {title} {\bibinfo {title} {First-principles study of twin boundaries and stacking faults in $\beta$-$\mathrm{Ga}_{2}\mathrm{O}_{3}$},\ }\href {https://doi.org/https://doi.org/10.1002/admi.202300318} {\bibfield  {journal} {\bibinfo  {journal} {Adv. Mater. Interfaces}\ ,\ \bibinfo {pages} {2300318}} (\bibinfo {year} {2023})}\BibitemShut {NoStop}%
\bibitem [{\citenamefont {Ngo}\ \emph {et~al.}(2020)\citenamefont {Ngo}, \citenamefont {Le}, \citenamefont {Lee}, \citenamefont {Hong}, \citenamefont {Ha}, \citenamefont {Lee},\ and\ \citenamefont {Moon}}]{2020homoepitaxial(-201)}%
  \BibitemOpen
  \bibfield  {author} {\bibinfo {author} {\bibfnamefont {T.~S.}\ \bibnamefont {Ngo}}, \bibinfo {author} {\bibfnamefont {D.~D.}\ \bibnamefont {Le}}, \bibinfo {author} {\bibfnamefont {J.}~\bibnamefont {Lee}}, \bibinfo {author} {\bibfnamefont {S.-K.}\ \bibnamefont {Hong}}, \bibinfo {author} {\bibfnamefont {J.-S.}\ \bibnamefont {Ha}}, \bibinfo {author} {\bibfnamefont {W.-S.}\ \bibnamefont {Lee}},\ and\ \bibinfo {author} {\bibfnamefont {Y.-B.}\ \bibnamefont {Moon}},\ }\bibfield  {title} {\bibinfo {title} {Investigation of defect structure in homoepitaxial \hkl(-201) $\beta$-\ce{Ga2O3} layers prepared by plasma-assisted molecular beam epitaxy},\ }\href {https://doi.org/https://doi.org/10.1016/j.jallcom.2020.155027} {\bibfield  {journal} {\bibinfo  {journal} {J. Alloys Compd.}\ }\textbf {\bibinfo {volume} {834}},\ \bibinfo {pages} {155027} (\bibinfo {year} {2020})}\BibitemShut {NoStop}%
\bibitem [{\citenamefont {Li}\ \emph {et~al.}(2020)\citenamefont {Li}, \citenamefont {Jiao}, \citenamefont {Yu}, \citenamefont {Hu}, \citenamefont {Lv}, \citenamefont {Li}, \citenamefont {Dong}, \citenamefont {Zhang}, \citenamefont {Zhang}, \citenamefont {Feng}, \citenamefont {Li},\ and\ \citenamefont {Du}}]{2021homoepitaxial(-201)}%
  \BibitemOpen
  \bibfield  {author} {\bibinfo {author} {\bibfnamefont {Z.}~\bibnamefont {Li}}, \bibinfo {author} {\bibfnamefont {T.}~\bibnamefont {Jiao}}, \bibinfo {author} {\bibfnamefont {J.}~\bibnamefont {Yu}}, \bibinfo {author} {\bibfnamefont {D.}~\bibnamefont {Hu}}, \bibinfo {author} {\bibfnamefont {Y.}~\bibnamefont {Lv}}, \bibinfo {author} {\bibfnamefont {W.}~\bibnamefont {Li}}, \bibinfo {author} {\bibfnamefont {X.}~\bibnamefont {Dong}}, \bibinfo {author} {\bibfnamefont {B.}~\bibnamefont {Zhang}}, \bibinfo {author} {\bibfnamefont {Y.}~\bibnamefont {Zhang}}, \bibinfo {author} {\bibfnamefont {Z.}~\bibnamefont {Feng}}, \bibinfo {author} {\bibfnamefont {G.}~\bibnamefont {Li}},\ and\ \bibinfo {author} {\bibfnamefont {G.}~\bibnamefont {Du}},\ }\bibfield  {title} {\bibinfo {title} {Single crystalline $\beta$-\ce{Ga2O3} homoepitaxial films grown by {MOCVD}},\ }\href {https://doi.org/https://doi.org/10.1016/j.vacuum.2020.109440} {\bibfield  {journal} {\bibinfo  {journal} {Vacuum}\ }\textbf {\bibinfo {volume} {178}},\ \bibinfo {pages} {109440} (\bibinfo {year} {2020})}\BibitemShut {NoStop}%
\bibitem [{\citenamefont {Polyakov}\ \emph {et~al.}(2021)\citenamefont {Polyakov}, \citenamefont {Smirnov}, \citenamefont {Shchemerov}, \citenamefont {Vasilev}, \citenamefont {Kochkova}, \citenamefont {Chernykh}, \citenamefont {Lagov}, \citenamefont {Pavlov}, \citenamefont {Stolbunov}, \citenamefont {Kulevoy}, \citenamefont {Borzykh}, \citenamefont {Lee}, \citenamefont {Ren},\ and\ \citenamefont {Pearton}}]{2021protonJAP}%
  \BibitemOpen
  \bibfield  {author} {\bibinfo {author} {\bibfnamefont {A.~Y.}\ \bibnamefont {Polyakov}}, \bibinfo {author} {\bibfnamefont {N.~B.}\ \bibnamefont {Smirnov}}, \bibinfo {author} {\bibfnamefont {I.~V.}\ \bibnamefont {Shchemerov}}, \bibinfo {author} {\bibfnamefont {A.~A.}\ \bibnamefont {Vasilev}}, \bibinfo {author} {\bibfnamefont {A.~I.}\ \bibnamefont {Kochkova}}, \bibinfo {author} {\bibfnamefont {A.~V.}\ \bibnamefont {Chernykh}}, \bibinfo {author} {\bibfnamefont {P.~B.}\ \bibnamefont {Lagov}}, \bibinfo {author} {\bibfnamefont {Y.~S.}\ \bibnamefont {Pavlov}}, \bibinfo {author} {\bibfnamefont {V.~S.}\ \bibnamefont {Stolbunov}}, \bibinfo {author} {\bibfnamefont {T.~V.}\ \bibnamefont {Kulevoy}}, \bibinfo {author} {\bibfnamefont {I.~V.}\ \bibnamefont {Borzykh}}, \bibinfo {author} {\bibfnamefont {I.-H.}\ \bibnamefont {Lee}}, \bibinfo {author} {\bibfnamefont {F.}~\bibnamefont {Ren}},\ and\ \bibinfo {author} {\bibfnamefont {S.~J.}\ \bibnamefont {Pearton}},\ }\bibfield  {title} {\bibinfo {title} {{Crystal orientation dependence of deep level spectra in proton irradiated bulk $\beta$-\ce{Ga2O3}}},\ }\href {https://doi.org/10.1063/5.0058555} {\bibfield  {journal} {\bibinfo  {journal} {J. Appl. Phys.}\ }\textbf {\bibinfo {volume} {130}},\ \bibinfo {pages} {035701} (\bibinfo {year} {2021})}\BibitemShut {NoStop}%
\bibitem [{\citenamefont {Tuttle}\ \emph {et~al.}(2023)\citenamefont {Tuttle}, \citenamefont {Karom}, \citenamefont {O’Hara}, \citenamefont {Schrimpf},\ and\ \citenamefont {Pantelides}}]{2023JAP}%
  \BibitemOpen
  \bibfield  {author} {\bibinfo {author} {\bibfnamefont {B.~R.}\ \bibnamefont {Tuttle}}, \bibinfo {author} {\bibfnamefont {N.~J.}\ \bibnamefont {Karom}}, \bibinfo {author} {\bibfnamefont {A.}~\bibnamefont {O’Hara}}, \bibinfo {author} {\bibfnamefont {R.~D.}\ \bibnamefont {Schrimpf}},\ and\ \bibinfo {author} {\bibfnamefont {S.~T.}\ \bibnamefont {Pantelides}},\ }\bibfield  {title} {\bibinfo {title} {{Atomic-displacement threshold energies and defect generation in irradiated $\beta$-$\mathrm{Ga}_{2}\mathrm{O}_{3}$: A first-principles investigation}},\ }\href {https://doi.org/10.1063/5.0124285} {\bibfield  {journal} {\bibinfo  {journal} {J. Appl. Phys.}\ }\textbf {\bibinfo {volume} {133}},\ \bibinfo {pages} {015703} (\bibinfo {year} {2023})}\BibitemShut {NoStop}%
\bibitem [{\citenamefont {He}\ \emph {et~al.}(2024)\citenamefont {He}, \citenamefont {Zhao}, \citenamefont {Byggmästar}, \citenamefont {He}, \citenamefont {Nordlund}, \citenamefont {He},\ and\ \citenamefont {Djurabekova}}]{he2024threshold}%
  \BibitemOpen
  \bibfield  {author} {\bibinfo {author} {\bibfnamefont {H.}~\bibnamefont {He}}, \bibinfo {author} {\bibfnamefont {J.}~\bibnamefont {Zhao}}, \bibinfo {author} {\bibfnamefont {J.}~\bibnamefont {Byggmästar}}, \bibinfo {author} {\bibfnamefont {R.}~\bibnamefont {He}}, \bibinfo {author} {\bibfnamefont {K.}~\bibnamefont {Nordlund}}, \bibinfo {author} {\bibfnamefont {C.}~\bibnamefont {He}},\ and\ \bibinfo {author} {\bibfnamefont {F.}~\bibnamefont {Djurabekova}},\ }\bibfield  {title} {\bibinfo {title} {Threshold displacement energy map of {Frenkel} pair generation in $\beta$-\ce{Ga2O3} from machine-learning-driven molecular dynamics simulations},\ }\href {https://doi.org/https://doi.org/10.1016/j.actamat.2024.120087} {\bibfield  {journal} {\bibinfo  {journal} {Acta Mater.}\ }\textbf {\bibinfo {volume} {276}},\ \bibinfo {pages} {120087} (\bibinfo {year} {2024})}\BibitemShut {NoStop}%
\bibitem [{\citenamefont {Thompson}\ \emph {et~al.}(2022)\citenamefont {Thompson}, \citenamefont {Aktulga}, \citenamefont {Berger}, \citenamefont {Bolintineanu}, \citenamefont {Brown}, \citenamefont {Crozier}, \citenamefont {{in 't Veld}}, \citenamefont {Kohlmeyer}, \citenamefont {Moore}, \citenamefont {Nguyen}, \citenamefont {Shan}, \citenamefont {Stevens}, \citenamefont {Tranchida}, \citenamefont {Trott},\ and\ \citenamefont {Plimpton}}]{lammps2022}%
  \BibitemOpen
  \bibfield  {author} {\bibinfo {author} {\bibfnamefont {A.~P.}\ \bibnamefont {Thompson}}, \bibinfo {author} {\bibfnamefont {H.~M.}\ \bibnamefont {Aktulga}}, \bibinfo {author} {\bibfnamefont {R.}~\bibnamefont {Berger}}, \bibinfo {author} {\bibfnamefont {D.~S.}\ \bibnamefont {Bolintineanu}}, \bibinfo {author} {\bibfnamefont {W.~M.}\ \bibnamefont {Brown}}, \bibinfo {author} {\bibfnamefont {P.~S.}\ \bibnamefont {Crozier}}, \bibinfo {author} {\bibfnamefont {P.~J.}\ \bibnamefont {{in 't Veld}}}, \bibinfo {author} {\bibfnamefont {A.}~\bibnamefont {Kohlmeyer}}, \bibinfo {author} {\bibfnamefont {S.~G.}\ \bibnamefont {Moore}}, \bibinfo {author} {\bibfnamefont {T.~D.}\ \bibnamefont {Nguyen}}, \bibinfo {author} {\bibfnamefont {R.}~\bibnamefont {Shan}}, \bibinfo {author} {\bibfnamefont {M.~J.}\ \bibnamefont {Stevens}}, \bibinfo {author} {\bibfnamefont {J.}~\bibnamefont {Tranchida}}, \bibinfo {author} {\bibfnamefont {C.}~\bibnamefont {Trott}},\ and\ \bibinfo {author} {\bibfnamefont {S.~J.}\ \bibnamefont {Plimpton}},\ }\bibfield  {title} {\bibinfo {title} {{LAMMPS} - a flexible simulation tool for particle-based materials modeling at the atomic, meso, and continuum scales},\ }\href {https://doi.org/10.1016/j.cpc.2021.108171} {\bibfield  {journal} {\bibinfo  {journal} {Comput. Phys. Commun.}\ }\textbf {\bibinfo {volume} {271}},\ \bibinfo {pages} {108171} (\bibinfo {year} {2022})}\BibitemShut {NoStop}%
\bibitem [{\citenamefont {Zhao}\ \emph {et~al.}(2023)\citenamefont {Zhao}, \citenamefont {Byggm{\"a}star}, \citenamefont {He}, \citenamefont {Nordlund}, \citenamefont {Djurabekova},\ and\ \citenamefont {Hua}}]{SFzhao2023complex}%
  \BibitemOpen
  \bibfield  {author} {\bibinfo {author} {\bibfnamefont {J.}~\bibnamefont {Zhao}}, \bibinfo {author} {\bibfnamefont {J.}~\bibnamefont {Byggm{\"a}star}}, \bibinfo {author} {\bibfnamefont {H.}~\bibnamefont {He}}, \bibinfo {author} {\bibfnamefont {K.}~\bibnamefont {Nordlund}}, \bibinfo {author} {\bibfnamefont {F.}~\bibnamefont {Djurabekova}},\ and\ \bibinfo {author} {\bibfnamefont {M.}~\bibnamefont {Hua}},\ }\bibfield  {title} {\bibinfo {title} {Complex $\mathrm{Ga}_{2}\mathrm{O}_{3}$ polymorphs explored by accurate and general-purpose machine-learning interatomic potentials},\ }\href {https://doi.org/10.1038/s41524-023-01117-1} {\bibfield  {journal} {\bibinfo  {journal} {npj Comput. Mater.}\ }\textbf {\bibinfo {volume} {9}},\ \bibinfo {pages} {159} (\bibinfo {year} {2023})}\BibitemShut {NoStop}%
\bibitem [{\citenamefont {Hoover}(1985)}]{hoover1985}%
  \BibitemOpen
  \bibfield  {author} {\bibinfo {author} {\bibfnamefont {W.~G.}\ \bibnamefont {Hoover}},\ }\bibfield  {title} {\bibinfo {title} {Canonical dynamics: Equilibrium phase-space distributions},\ }\href {https://doi.org/10.1103/PhysRevA.31.1695} {\bibfield  {journal} {\bibinfo  {journal} {Phys. Rev. A}\ }\textbf {\bibinfo {volume} {31}},\ \bibinfo {pages} {1695} (\bibinfo {year} {1985})}\BibitemShut {NoStop}%
\bibitem [{\citenamefont {Nordlund}\ \emph {et~al.}(2016)\citenamefont {Nordlund}, \citenamefont {Djurabekova},\ and\ \citenamefont {Hobler}}]{Nordlund-PhysRevB.94.214109}%
  \BibitemOpen
  \bibfield  {author} {\bibinfo {author} {\bibfnamefont {K.}~\bibnamefont {Nordlund}}, \bibinfo {author} {\bibfnamefont {F.}~\bibnamefont {Djurabekova}},\ and\ \bibinfo {author} {\bibfnamefont {G.}~\bibnamefont {Hobler}},\ }\bibfield  {title} {\bibinfo {title} {Large fraction of crystal directions leads to ion channeling},\ }\href {https://doi.org/10.1103/PhysRevB.94.214109} {\bibfield  {journal} {\bibinfo  {journal} {Phys. Rev. B}\ }\textbf {\bibinfo {volume} {94}},\ \bibinfo {pages} {214109} (\bibinfo {year} {2016})}\BibitemShut {NoStop}%
\bibitem [{\citenamefont {Nordlund}\ \emph {et~al.}(1998)\citenamefont {Nordlund}, \citenamefont {Ghaly}, \citenamefont {Averback}, \citenamefont {Caturla}, \citenamefont {Diaz de~la Rubia},\ and\ \citenamefont {Tarus}}]{nordlund1998defect}%
  \BibitemOpen
  \bibfield  {author} {\bibinfo {author} {\bibfnamefont {K.}~\bibnamefont {Nordlund}}, \bibinfo {author} {\bibfnamefont {M.}~\bibnamefont {Ghaly}}, \bibinfo {author} {\bibfnamefont {R.~S.}\ \bibnamefont {Averback}}, \bibinfo {author} {\bibfnamefont {M.}~\bibnamefont {Caturla}}, \bibinfo {author} {\bibfnamefont {T.}~\bibnamefont {Diaz de~la Rubia}},\ and\ \bibinfo {author} {\bibfnamefont {J.}~\bibnamefont {Tarus}},\ }\bibfield  {title} {\bibinfo {title} {Defect production in collision cascades in elemental semiconductors and fcc metals},\ }\href {https://doi.org/10.1103/PhysRevB.57.7556} {\bibfield  {journal} {\bibinfo  {journal} {Phys. Rev. B}\ }\textbf {\bibinfo {volume} {57}},\ \bibinfo {pages} {7556} (\bibinfo {year} {1998})}\BibitemShut {NoStop}%
\bibitem [{\citenamefont {Stukowski}(2010)}]{ovito2010}%
  \BibitemOpen
  \bibfield  {author} {\bibinfo {author} {\bibfnamefont {A.}~\bibnamefont {Stukowski}},\ }\bibfield  {title} {\bibinfo {title} {Visualization and analysis of atomistic simulation data with {OVITO}–the open visualization tool},\ }\href {https://doi.org/10.1088/0965-0393/18/1/015012} {\bibfield  {journal} {\bibinfo  {journal} {Modell. Simul. Mater. Sci. Eng.}\ }\textbf {\bibinfo {volume} {18}},\ \bibinfo {pages} {015012} (\bibinfo {year} {2010})}\BibitemShut {NoStop}%
\bibitem [{\citenamefont {Kresse}\ and\ \citenamefont {Furthm\"uller}(1996)}]{PhysRevB.54.11169}%
  \BibitemOpen
  \bibfield  {author} {\bibinfo {author} {\bibfnamefont {G.}~\bibnamefont {Kresse}}\ and\ \bibinfo {author} {\bibfnamefont {J.}~\bibnamefont {Furthm\"uller}},\ }\bibfield  {title} {\bibinfo {title} {Efficient iterative schemes for ab initio total-energy calculations using a plane-wave basis set},\ }\href {https://doi.org/10.1103/PhysRevB.54.11169} {\bibfield  {journal} {\bibinfo  {journal} {Phys. Rev. B}\ }\textbf {\bibinfo {volume} {54}},\ \bibinfo {pages} {11169} (\bibinfo {year} {1996})}\BibitemShut {NoStop}%
\bibitem [{\citenamefont {Bl\"ochl}(1994)}]{PhysRevB.50.17953_PAW}%
  \BibitemOpen
  \bibfield  {author} {\bibinfo {author} {\bibfnamefont {P.~E.}\ \bibnamefont {Bl\"ochl}},\ }\bibfield  {title} {\bibinfo {title} {Projector augmented-wave method},\ }\href {https://doi.org/10.1103/PhysRevB.50.17953} {\bibfield  {journal} {\bibinfo  {journal} {Phys. Rev. B}\ }\textbf {\bibinfo {volume} {50}},\ \bibinfo {pages} {17953} (\bibinfo {year} {1994})}\BibitemShut {NoStop}%
\bibitem [{\citenamefont {Perdew}\ \emph {et~al.}(1996)\citenamefont {Perdew}, \citenamefont {Burke},\ and\ \citenamefont {Ernzerhof}}]{PhysRevLett.77.3865}%
  \BibitemOpen
  \bibfield  {author} {\bibinfo {author} {\bibfnamefont {J.~P.}\ \bibnamefont {Perdew}}, \bibinfo {author} {\bibfnamefont {K.}~\bibnamefont {Burke}},\ and\ \bibinfo {author} {\bibfnamefont {M.}~\bibnamefont {Ernzerhof}},\ }\bibfield  {title} {\bibinfo {title} {Generalized gradient approximation made simple},\ }\href {https://doi.org/10.1103/PhysRevLett.77.3865} {\bibfield  {journal} {\bibinfo  {journal} {Phys. Rev. Lett.}\ }\textbf {\bibinfo {volume} {77}},\ \bibinfo {pages} {3865} (\bibinfo {year} {1996})}\BibitemShut {NoStop}%
\bibitem [{\citenamefont {{Heyd}}\ \emph {et~al.}(2006)\citenamefont {{Heyd}}, \citenamefont {{Scuseria}},\ and\ \citenamefont {{Ernzerhof}}}]{2006JChPh.124u9906H_HSE}%
  \BibitemOpen
  \bibfield  {author} {\bibinfo {author} {\bibfnamefont {J.}~\bibnamefont {{Heyd}}}, \bibinfo {author} {\bibfnamefont {G.~E.}\ \bibnamefont {{Scuseria}}},\ and\ \bibinfo {author} {\bibfnamefont {M.}~\bibnamefont {{Ernzerhof}}},\ }\bibfield  {title} {\bibinfo {title} {{Erratum: ``Hybrid functionals based on a screened Coulomb potential'' [J. Chem. Phys. 118, 8207 (2003)]}},\ }\href {https://doi.org/10.1063/1.2204597} {\bibfield  {journal} {\bibinfo  {journal} {\jcp}\ }\textbf {\bibinfo {volume} {124}},\ \bibinfo {pages} {219906} (\bibinfo {year} {2006})}\BibitemShut {NoStop}%
\bibitem [{\citenamefont {Varley}(2020)}]{Varley2020_book}%
  \BibitemOpen
  \bibfield  {author} {\bibinfo {author} {\bibfnamefont {J.~B.}\ \bibnamefont {Varley}},\ }\bibinfo {title} {First-principles calculations 2: Doping and defects in {$\rm Ga_2O_3$}},\ in\ \href {https://doi.org/10.1007/978-3-030-37153-1_18} {\emph {\bibinfo {booktitle} {Gallium Oxide: Materials Properties, Crystal Growth, and Devices}}},\ \bibinfo {editor} {edited by\ \bibinfo {editor} {\bibfnamefont {M.}~\bibnamefont {Higashiwaki}}\ and\ \bibinfo {editor} {\bibfnamefont {S.}~\bibnamefont {Fujita}}}\ (\bibinfo  {publisher} {Springer International Publishing},\ \bibinfo {year} {2020})\ pp.\ \bibinfo {pages} {329--348}\BibitemShut {NoStop}%
\bibitem [{\citenamefont {Huang}\ \emph {et~al.}(2022)\citenamefont {Huang}, \citenamefont {Zheng}, \citenamefont {Dai}, \citenamefont {Guo}, \citenamefont {Wang}, \citenamefont {Jiang}, \citenamefont {Wei},\ and\ \citenamefont {Chen}}]{Huang_2022}%
  \BibitemOpen
  \bibfield  {author} {\bibinfo {author} {\bibfnamefont {M.}~\bibnamefont {Huang}}, \bibinfo {author} {\bibfnamefont {Z.}~\bibnamefont {Zheng}}, \bibinfo {author} {\bibfnamefont {Z.}~\bibnamefont {Dai}}, \bibinfo {author} {\bibfnamefont {X.}~\bibnamefont {Guo}}, \bibinfo {author} {\bibfnamefont {S.}~\bibnamefont {Wang}}, \bibinfo {author} {\bibfnamefont {L.}~\bibnamefont {Jiang}}, \bibinfo {author} {\bibfnamefont {J.}~\bibnamefont {Wei}},\ and\ \bibinfo {author} {\bibfnamefont {S.}~\bibnamefont {Chen}},\ }\bibfield  {title} {\bibinfo {title} {Dasp: Defect and dopant ab-initio simulation package},\ }\href {https://doi.org/10.1088/1674-4926/43/4/042101} {\bibfield  {journal} {\bibinfo  {journal} {J. Semicond.}\ }\textbf {\bibinfo {volume} {43}},\ \bibinfo {pages} {042101} (\bibinfo {year} {2022})}\BibitemShut {NoStop}%
\bibitem [{\citenamefont {Van~de Walle}\ and\ \citenamefont {Neugebauer}(2004)}]{2004vdwreview}%
  \BibitemOpen
  \bibfield  {author} {\bibinfo {author} {\bibfnamefont {C.~G.}\ \bibnamefont {Van~de Walle}}\ and\ \bibinfo {author} {\bibfnamefont {J.}~\bibnamefont {Neugebauer}},\ }\bibfield  {title} {\bibinfo {title} {{First-principles calculations for defects and impurities: Applications to III-nitrides}},\ }\href {https://doi.org/10.1063/1.1682673} {\bibfield  {journal} {\bibinfo  {journal} {J. Appl. Phys.}\ }\textbf {\bibinfo {volume} {95}},\ \bibinfo {pages} {3851} (\bibinfo {year} {2004})}\BibitemShut {NoStop}%
\bibitem [{\citenamefont {Freysoldt}\ \emph {et~al.}(2009)\citenamefont {Freysoldt}, \citenamefont {Neugebauer},\ and\ \citenamefont {Van~de Walle}}]{PhysRevLett.102.016402}%
  \BibitemOpen
  \bibfield  {author} {\bibinfo {author} {\bibfnamefont {C.}~\bibnamefont {Freysoldt}}, \bibinfo {author} {\bibfnamefont {J.}~\bibnamefont {Neugebauer}},\ and\ \bibinfo {author} {\bibfnamefont {C.~G.}\ \bibnamefont {Van~de Walle}},\ }\bibfield  {title} {\bibinfo {title} {Fully ab initio finite-size corrections for charged-defect supercell calculations},\ }\href {https://doi.org/10.1103/PhysRevLett.102.016402} {\bibfield  {journal} {\bibinfo  {journal} {Phys. Rev. Lett.}\ }\textbf {\bibinfo {volume} {102}},\ \bibinfo {pages} {016402} (\bibinfo {year} {2009})}\BibitemShut {NoStop}%
\bibitem [{\citenamefont {Freysoldt}\ \emph {et~al.}(2011)\citenamefont {Freysoldt}, \citenamefont {Neugebauer},\ and\ \citenamefont {Van~de Walle}}]{FNV2010}%
  \BibitemOpen
  \bibfield  {author} {\bibinfo {author} {\bibfnamefont {C.}~\bibnamefont {Freysoldt}}, \bibinfo {author} {\bibfnamefont {J.}~\bibnamefont {Neugebauer}},\ and\ \bibinfo {author} {\bibfnamefont {C.~G.}\ \bibnamefont {Van~de Walle}},\ }\bibfield  {title} {\bibinfo {title} {Electrostatic interactions between charged defects in supercells},\ }\href {https://doi.org/10.1002/pssb.201046289} {\bibfield  {journal} {\bibinfo  {journal} {Phys. Status Solidi B}\ }\textbf {\bibinfo {volume} {248}},\ \bibinfo {pages} {1067} (\bibinfo {year} {2011})}\BibitemShut {NoStop}%
\bibitem [{\citenamefont {Huang}\ \emph {et~al.}(2021)\citenamefont {Huang}, \citenamefont {Cai}, \citenamefont {Wang}, \citenamefont {Gong}, \citenamefont {Wei},\ and\ \citenamefont {Chen}}]{2021HuangSmall}%
  \BibitemOpen
  \bibfield  {author} {\bibinfo {author} {\bibfnamefont {M.}~\bibnamefont {Huang}}, \bibinfo {author} {\bibfnamefont {Z.}~\bibnamefont {Cai}}, \bibinfo {author} {\bibfnamefont {S.}~\bibnamefont {Wang}}, \bibinfo {author} {\bibfnamefont {X.}~\bibnamefont {Gong}}, \bibinfo {author} {\bibfnamefont {S.}~\bibnamefont {Wei}},\ and\ \bibinfo {author} {\bibfnamefont {S.}~\bibnamefont {Chen}},\ }\bibfield  {title} {\bibinfo {title} {More se vacancies in \ce{Sb2Se3} under se-rich conditions: An abnormal behavior induced by defect-correlation in compensated compound semiconductors},\ }\href {https://doi.org/https://doi.org/10.1002/smll.202102429} {\bibfield  {journal} {\bibinfo  {journal} {Small}\ }\textbf {\bibinfo {volume} {17}},\ \bibinfo {pages} {2102429} (\bibinfo {year} {2021})}\BibitemShut {NoStop}%
\bibitem [{\citenamefont {Ma}\ \emph {et~al.}(2011)\citenamefont {Ma}, \citenamefont {Wei}, \citenamefont {Gessert},\ and\ \citenamefont {Chin}}]{PhysRevB.83.245207_degeneracy_factor}%
  \BibitemOpen
  \bibfield  {author} {\bibinfo {author} {\bibfnamefont {J.}~\bibnamefont {Ma}}, \bibinfo {author} {\bibfnamefont {S.}~\bibnamefont {Wei}}, \bibinfo {author} {\bibfnamefont {T.~A.}\ \bibnamefont {Gessert}},\ and\ \bibinfo {author} {\bibfnamefont {K.~K.}\ \bibnamefont {Chin}},\ }\bibfield  {title} {\bibinfo {title} {Carrier density and compensation in semiconductors with multiple dopants and multiple transition energy levels: Case of {Cu} impurities in {CdTe}},\ }\href {https://doi.org/10.1103/PhysRevB.83.245207} {\bibfield  {journal} {\bibinfo  {journal} {Phys. Rev. B}\ }\textbf {\bibinfo {volume} {83}},\ \bibinfo {pages} {245207} (\bibinfo {year} {2011})}\BibitemShut {NoStop}%
\bibitem [{\citenamefont {Li}\ \emph {et~al.}(2019)\citenamefont {Li}, \citenamefont {Zhang}, \citenamefont {Lyu},\ and\ \citenamefont {Li}}]{2019highT}%
  \BibitemOpen
  \bibfield  {author} {\bibinfo {author} {\bibfnamefont {M.}~\bibnamefont {Li}}, \bibinfo {author} {\bibfnamefont {L.}~\bibnamefont {Zhang}}, \bibinfo {author} {\bibfnamefont {S.}~\bibnamefont {Lyu}},\ and\ \bibinfo {author} {\bibfnamefont {Z.}~\bibnamefont {Li}},\ }\bibfield  {title} {\bibinfo {title} {Effects of ion irradiation and oxidation on point defects in {IG-110} nuclear grade graphite},\ }\href {https://doi.org/10.7498/aps.68.20190371} {\bibfield  {journal} {\bibinfo  {journal} {Acta Phys. Sin.}\ }\textbf {\bibinfo {volume} {68}},\ \bibinfo {pages} {128102} (\bibinfo {year} {2019})}\BibitemShut {NoStop}%
\bibitem [{\citenamefont {Zhao}\ \emph {et~al.}(2024)\citenamefont {Zhao}, \citenamefont {Garc{\'i}a~Fern{\'a}ndez}, \citenamefont {Azarov}, \citenamefont {He}, \citenamefont {Prytz}, \citenamefont {Nordlund}, \citenamefont {Hua}, \citenamefont {Djurabekova},\ and\ \citenamefont {Kuznetsov}}]{SFzhao2024crystallization}%
  \BibitemOpen
  \bibfield  {author} {\bibinfo {author} {\bibfnamefont {J.}~\bibnamefont {Zhao}}, \bibinfo {author} {\bibfnamefont {J.}~\bibnamefont {Garc{\'i}a~Fern{\'a}ndez}}, \bibinfo {author} {\bibfnamefont {A.}~\bibnamefont {Azarov}}, \bibinfo {author} {\bibfnamefont {R.}~\bibnamefont {He}}, \bibinfo {author} {\bibfnamefont {{\O}.}~\bibnamefont {Prytz}}, \bibinfo {author} {\bibfnamefont {K.}~\bibnamefont {Nordlund}}, \bibinfo {author} {\bibfnamefont {M.}~\bibnamefont {Hua}}, \bibinfo {author} {\bibfnamefont {F.}~\bibnamefont {Djurabekova}},\ and\ \bibinfo {author} {\bibfnamefont {A.}~\bibnamefont {Kuznetsov}},\ }\bibfield  {title} {\bibinfo {title} {Crystallization instead of amorphization in collision cascades in gallium oxide},\ }\bibfield  {journal} {\bibinfo  {journal} {arXiv}\ }\href {https://doi.org/10.48550/arXiv.2401.07675} {10.48550/arXiv.2401.07675} (\bibinfo {year} {2024})\BibitemShut {NoStop}%
\bibitem [{\citenamefont {Huang}\ \emph {et~al.}(2023{\natexlab{b}})\citenamefont {Huang}, \citenamefont {Xu}, \citenamefont {Yang}, \citenamefont {Yu}, \citenamefont {Wei}, \citenamefont {Ying}, \citenamefont {Liu}, \citenamefont {Jing}, \citenamefont {Li},\ and\ \citenamefont {Li}}]{HUANG2023_HSE_formationenergy}%
  \BibitemOpen
  \bibfield  {author} {\bibinfo {author} {\bibfnamefont {Y.}~\bibnamefont {Huang}}, \bibinfo {author} {\bibfnamefont {X.}~\bibnamefont {Xu}}, \bibinfo {author} {\bibfnamefont {J.}~\bibnamefont {Yang}}, \bibinfo {author} {\bibfnamefont {X.}~\bibnamefont {Yu}}, \bibinfo {author} {\bibfnamefont {Y.}~\bibnamefont {Wei}}, \bibinfo {author} {\bibfnamefont {T.}~\bibnamefont {Ying}}, \bibinfo {author} {\bibfnamefont {Z.}~\bibnamefont {Liu}}, \bibinfo {author} {\bibfnamefont {Y.}~\bibnamefont {Jing}}, \bibinfo {author} {\bibfnamefont {W.}~\bibnamefont {Li}},\ and\ \bibinfo {author} {\bibfnamefont {X.}~\bibnamefont {Li}},\ }\bibfield  {title} {\bibinfo {title} {Library of intrinsic defects in $\beta$-$\mathrm{Ga}_{2}\mathrm{O}_{3}$: First-principles studies},\ }\href {https://doi.org/https://doi.org/10.1016/j.mtcomm.2023.105898} {\bibfield  {journal} {\bibinfo  {journal} {Mater. Today Commun.}\ }\textbf {\bibinfo {volume} {35}},\ \bibinfo {pages} {105898} (\bibinfo {year} {2023}{\natexlab{b}})}\BibitemShut {NoStop}%
\bibitem [{\citenamefont {von Bardeleben}\ \emph {et~al.}(2019)\citenamefont {von Bardeleben}, \citenamefont {Zhou}, \citenamefont {Gerstmann}, \citenamefont {Skachkov}, \citenamefont {Lambrecht}, \citenamefont {Ho},\ and\ \citenamefont {Deák}}]{2018ProtonII}%
  \BibitemOpen
  \bibfield  {author} {\bibinfo {author} {\bibfnamefont {H.~J.}\ \bibnamefont {von Bardeleben}}, \bibinfo {author} {\bibfnamefont {S.}~\bibnamefont {Zhou}}, \bibinfo {author} {\bibfnamefont {U.}~\bibnamefont {Gerstmann}}, \bibinfo {author} {\bibfnamefont {D.}~\bibnamefont {Skachkov}}, \bibinfo {author} {\bibfnamefont {W.~R.~L.}\ \bibnamefont {Lambrecht}}, \bibinfo {author} {\bibfnamefont {Q.~D.}\ \bibnamefont {Ho}},\ and\ \bibinfo {author} {\bibfnamefont {P.}~\bibnamefont {Deák}},\ }\bibfield  {title} {\bibinfo {title} {{Proton irradiation induced defects in $\beta$-\ce{Ga2O3}: A combined EPR and theory study}},\ }\href {https://doi.org/10.1063/1.5053158} {\bibfield  {journal} {\bibinfo  {journal} {APL Mater.}\ }\textbf {\bibinfo {volume} {7}},\ \bibinfo {pages} {022521} (\bibinfo {year} {2019})}\BibitemShut {NoStop}%
\end{thebibliography}%

\end{document}